\documentclass[journal]{IEEEtran}
\usepackage{cite}
\usepackage{amsmath,amssymb,amsfonts}
\usepackage{algorithmic}
\usepackage{graphicx}
\usepackage{textcomp}
\usepackage{xcolor}
\usepackage{cuted}
\usepackage{epstopdf}
\usepackage{stfloats}
\usepackage{algorithm}
\usepackage{enumerate}

\usepackage[caption=false,font=small,labelfont=sf,textfont=sf]{subfig}

\newenvironment{proof}{{\indent \indent \it Proof:\,\,}}{\hfill $\blacksquare$\par}

\begin{document}

\title{Two-Timescale Design for Movable Antenna-Enabled Multiuser MIMO Systems}
\author{Ziyuan~Zheng, Qingqing~Wu, Wen~Chen, and
        Guojie~Hu     \vspace{-36pt}
\thanks{Z. Zheng, Q. Wu, and W. Chen are with the Department of Electronic Engineering, Shanghai Jiao Tong University, 200240, China (e-mail: zhengziyuan2024@sjtu.edu.cn, qingqingwu@sjtu.edu.cn, wenchen@sjtu.edu.cn).; G. Hu is with the College of Communication Engineering, Rocket Force University of Engineering, Xi’an, 710025, China (e-mail: lgdxhgj@sina.com).}
\thanks{This work was supported by National Key R\&D Program of China (2022YFB2903500), Guangdong science and technology program under grant 2022A0505050011, and NSFC 62371289 and NSFC 62331022.}
}

\markboth{}%
{Shell \MakeLowercase{\textit{et al.}}: Bare Demo of IEEEtran.cls for IEEE Journals}

\maketitle

\begin{abstract}
Movable antennas (MAs), which can be swiftly repositioned within a defined region, offer a promising solution to the limitations of fixed-position antennas (FPAs) in adapting to spatial variations in wireless channels, thereby improving channel conditions and communication between transceivers. However, frequent MA position adjustments based on instantaneous channel state information (CSI) incur high operational complexity, making real-time CSI acquisition impractical, especially in fast-fading channels.
To address these challenges, we propose a two-timescale transmission framework for MA-enabled multiuser multiple-input-multiple-output (MU-MIMO) systems. In the large timescale, statistical CSI is exploited to optimize MA positions for long-term ergodic performance, whereas, in the small timescale, beamforming vectors are designed using instantaneous CSI to handle short-term channel fluctuations. Within this new framework, we analyze the ergodic sum rate and develop efficient MA position optimization algorithms for both maximum-ratio-transmission (MRT) and zero-forcing (ZF) beamforming schemes. These algorithms employ alternating optimization (AO), successive convex approximation (SCA), and majorization-minimization (MM) techniques, iteratively optimizing antenna positions and refining surrogate functions that approximate the ergodic sum rate.
Numerical results show significant ergodic sum rate gains with the proposed two-timescale MA design over conventional FPA systems, particularly under moderate to strong line-of-sight (LoS) conditions. Notably, MA with ZF beamforming consistently outperforms MA with MRT, highlighting the synergy between beamforming and MAs for superior interference management in environments with moderate Rician factors and high user density, while MA with MRT can offer a simplified alternative to complex beamforming designs in strong LoS conditions. 
\end{abstract}

\vspace{-5pt}

\IEEEpeerreviewmaketitle

\vspace{-12pt}
\section{Introduction}
\vspace{-6pt}
Multiple-input multiple-output (MIMO) has been established as a critical enabler for enhancing capacity, reliability, and overall performance in the evolution of wireless communication systems [1]. MIMO systems provide significant gains in beamforming, spatial multiplexing, and diversity by leveraging independent or quasi-independent channel fading. These gains enable the simultaneous transmission of multiple data streams, significantly boosting spectral efficiency compared to single-antenna systems [2], [3]. As wireless networks transition to higher frequency bands such as millimeter-wave and terahertz, massive MIMO has become crucial for mitigating propagation loss through large-scale antenna arrays and achieving near-orthogonal user channels, thereby improving spectral efficiency and link reliability significantly [4], [5], [6]. Compared to traditional MIMO, massive MIMO's ability to harness spatial correlation for array gains and interference mitigation renders it essential for next-generation wireless systems.

However, deploying large-scale antenna arrays introduces several challenges, particularly for escalating hardware costs and increased radio-frequency (RF) energy consumption, which hinder the development of sustainable and energy-efficient networks [7], [8]. With the advancement towards 6G networks, the limitations of traditional fixed-position antennas (FPAs) become increasingly evident. Due to their stationary nature, FPAs cannot fully leverage spatial variations in wireless channels. To compensate, conventional MIMO and massive MIMO systems resort to increasing the number of antennas and RF chains. While this approach enhances service quality, it comes at the expense of increased energy consumption and hardware cost. Fundamentally, the static nature of FPAs renders them less adaptable to dynamic channel conditions and user distributions, leading to suboptimal performance and constraining their ability to fully exploit spatial diversity and multiplexing [9], [10].

Movable antennas (MAs) [11], also referred to as fluid antennas [12], overcome the limitations of FPAs by enabling flexible positioning within a spatial region. Connected to the RF chain via flexible cables and controlled by drivers, MAs can be swiftly positioned at locations with favorable channel conditions to reshape wireless channels and optimize communication between transceivers. This flexibility allows MAs to exploit spatial degrees of freedom, avoiding deep fading or interference-prone areas without the need for additional antennas or RF chains. By dynamically adjusting antenna positions in response to channel conditions and user distribution, MAs can fully exploit spatial diversity to maximize channel gain, thereby enhancing signal-to-noise ratio (SNR) [13]. MAs also enhance interference mitigation by moving to locations with deep fading relative to interference sources, boosting signal-to-interference-plus-noise ratio (SINR) without requiring multiple antennas [14]. Additionally, MA-enabled MIMO systems optimize spatial multiplexing rates by dynamically adjusting antenna positions to reshape the channel matrix, thus maximizing MIMO capacity [15].

The inherent advantages of MAs make them a promising candidate for integration into 6G networks, sparking significant interest in their deployment across various wireless communication systems, particularly in optimizing antenna positioning.
Early research primarily focused on point-to-point single-user systems. In [14], a mechanical MA architecture was proposed within a single-input single-output (SISO) system to enhance SNR gains, which were found to be highly dependent on the number of channel paths and the movement range of the antennas. This concept was later extended to point-to-point MIMO systems in [15], where an alternating optimization (AO) algorithm was employed to jointly optimize the positions of MAs and the transmit signal covariance, thereby maximizing channel capacity.
Besides point-to-point systems, the application of MAs has been investigated in various multiuser scenarios [16], [17]. Several suboptimal algorithms have been proposed for uplink transmissions involving multiple single-MA users communicating with a base station (BS) equipped with an FPA array. Projected gradient descent techniques were utilized to minimize user transmit power while guaranteeing quality of service [18]. Moreover, particle swarm optimization was applied to improve user fairness for adjusting MA positions and users' transmit power, showcasing the capability of MAs to manage multiuser interference effectively [19].
Downlink scenarios have also been investigated [20]-[25]. For example, integrating sub-connected MAs with hybrid beamforming schemes outperformed fully connected FPA arrays in terms of achievable sum rates under certain conditions [20]. This integration, alongside other studies [14], demonstrated that even minor movements of MAs (within sub-wavelength distances) can substantially alter channel conditions due to small-scale fading, leading to enhanced communication performance.
Recent research further indicates that MAs can be effectively applied in multi-cell MISO [22], multi-group multicast [23], full-duplex communications [24], and wideband systems [25]. 

Despite the promising potential of MA technology, research in this field remains in its early stage, with most studies focusing on algorithmic design and theoretical performance limits. These studies generally assume that each antenna element can be adjusted globally and in real-time within the transmit/receive region, relying on the availability of instantaneous channel state information (CSI) [26]. However, such an approach presents several critical challenges.
First, although mechanical MA systems, as discussed in [16] and [17], provide flexible movement in 2D and 3D spaces, this flexibility incurs substantial movement energy consumption due to the need for frequent adjustments to track optimal positions, compounded by delays resulting from mechanical response speed limitations. Fluid antenna systems (FAS) can be more compact and energy-efficient [12], [13], [27]. 
Second, the reliance on complete and instantaneous CSI across the movement region to optimize MA positions poses an additional fundamental challenge. Most existing schemes rely on full CSI; however, acquiring accurate real-time CSI across the entire movement region is impractical and even infeasible, especially in fast-fading channels. This limitation hinders the practical deployment of MAs in real-world systems.


To address these challenges, an approach that involves adjusting antenna positions on a longer timescale is proposed by utilizing statistical CSI, such as user signal power, user distribution, and large-scale channel parameters, in accordance with current network protocols [28]. This method reduces the overhead and complexity associated with channel estimation while still providing performance gains over traditional FPAs. Another similar research explored scenarios where only statistical CSI is available, proposing simplified antenna movement modes and optimization frameworks that focus on transmit precoding and antenna position design [29].  
However, relying exclusively on statistical CSI can result in performance degradation [30], [31], as it fails to account for rapidly changing channel conditions, especially in multi-path or highly dynamic environments. Therefore, it is imperative to develop a general optimization framework that balances performance improvements with practical implementation challenges in MA-enabled systems.

In this paper, we propose a novel two-timescale transmission framework for MA-enabled multiuser MIMO (MU-MIMO) systems. The key contributions are summarized as follows:
\vspace{-15pt}
\begin{enumerate}
\item We propose a two-timescale transmission scheme for MA-enabled systems. The large-timescale optimization exploits statistical CSI to determine optimal MA positions, maximizing long-term ergodic performance, while in the small timescale, beamforming vectors are derived from instantaneous CSI to adapt to short-term channel fluctuations. This decoupling of MA position optimization from the instantaneous transmission process provides a solution that strikes a balance between performance and practicality, reducing the frequency of MA updates and lowering channel estimation overhead.
\item Within the proposed two-timescale framework, we analyze the ergodic sum rate under maximum-ratio-transmission (MRT) beamforming and develop an efficient antenna position optimization algorithm. This algorithm employs alternating optimization (AO) and successive convex approximation (SCA) techniques to iteratively optimize MAs' positions in an element-wise manner, enhancing the system's ergodic performance.
\item We further analyze zero-forcing (ZF) beamforming and propose an algorithm that resorts to AO, SCA, and majorization-minimization (MM) techniques. This iterative approach optimizes a surrogate function, which serves as a lower bound for the ergodic sum rate, to refine the antenna positions for enhanced performance. In addition, we incorporate the effect of spatial correlation into the channel model and analyze its impact on the ergodic rate for both cases. 
\item Numerical results verify the superiority of the proposed two-timescale MA design compared to conventional FPA systems, indicating considerable ergodic sum rate improvements under moderate to strong line-of-sight (LoS) conditions. Notably, MA combined with ZF beamforming consistently outperforms MA with MRT beamforming, underscoring the synergy between beamforming and MA techniques for superior interference management, especially in environments with moderate Rician factors and high user density. In contrast, MA with MRT beamforming offers a simplified alternative to sophisticated beamforming designs in strong LoS conditions.
\end{enumerate}
\vspace{-3pt}
It is worth emphasizing that the proposed two-timescale design offers distinct advantages across different deployment scenarios. In semi-static environments such as industrial IoT, smart home, and fixed wireless access, the framework achieves substantial and stable performance gains with dramatically reduced CSI overhead and minimal MA updates, making it highly practical and effective. For mobile users, although frequent statistical CSI and MA position updates may be required to maintain performance, the design still effectively reduces instantaneous CSI overhead and provides a flexible fallback to FPA-based configurations when positioning inaccuracies become significant. This adaptability ensures the practicality of the proposed scheme under a wide range of network conditions.

The rest of the paper is structured as follows: Section II introduces the system model and problem formulation. Section III presents the achievable rate derivation and antenna position optimization under MRT beamforming. Section IV derives the achievable rate and optimizes MAs' positions under ZF beamforming. Section V derives the achievable rate with spatial correlation. Section VI presents the numerical results, and Section VII concludes the paper.

\textit{Notations:} Scalars are denoted by italic letters, and vectors and matrices are denoted by bold-face lower and upper-case letters, respectively. $\mathbb{C} ^{x\times y}$ denotes the space of complex matrices. $\left| x \right|$ denotes the modulus of a complex-valued scalar $x$. For a vector $\boldsymbol{x}$, $\left\| \boldsymbol{x} \right\|$ denotes its Euclidean norm and $\boldsymbol{x}^\dagger$ denotes its conjugate. The distribution of a circularly symmetric complex Gaussian random vector with mean vector $\boldsymbol{x}$ and covariance matrix $\boldsymbol{\varSigma }$ is denoted by $\mathcal{CN} \left( \boldsymbol{x},\boldsymbol{\varSigma } \right)$. The Euclidean gradient of a scalar function $f(\boldsymbol{x})$ with a vector variable $\boldsymbol{x}$ is denoted by $\nabla f(\boldsymbol{x})$. For a square matrix $\boldsymbol{S}$, $\mathrm{tr}\left( \boldsymbol{S} \right)$ denote its trace. For any general matrix, $\boldsymbol{M}^H$ and $\left[ \boldsymbol{M} \right] _{ij}$ denote its conjugate transpose and $(i,j)$th element,respectively. $\boldsymbol{I}$ denotes an identity matrix, respectively. $\mathbb{E}[\cdot]$ stands for the expectation operation.

\setlength{\abovecaptionskip}{-6pt}
 \begin{figure}
\centering
\includegraphics[width=3.4in,height=1.31in]{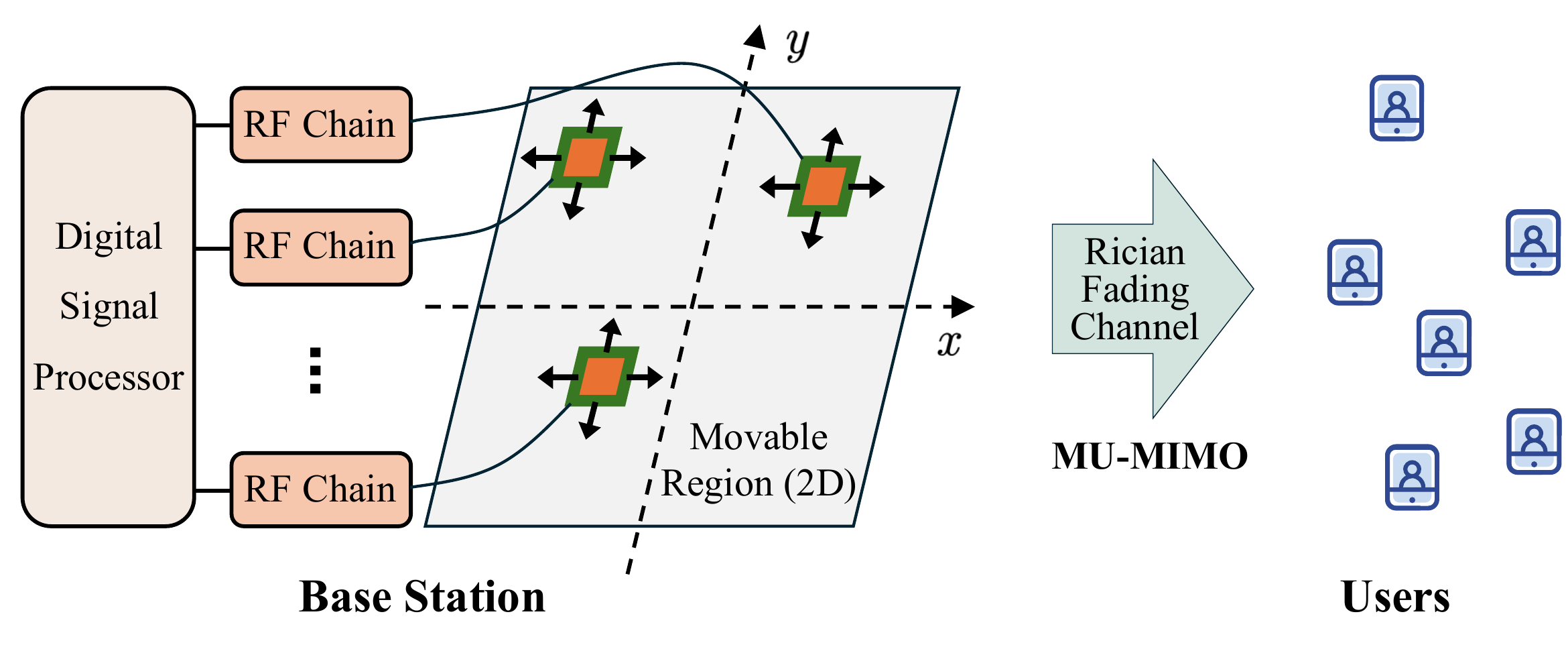}
\captionsetup{font=small}
\caption{Illustration of an MA-enabled multiuser system.} \label{fig1}
\vspace{-18pt}
\end{figure}

\vspace{-9pt}
\section{System Model and Problem Formulation}
\subsection{MA-enabled MU-MIMO System}
\vspace{-3pt}
As shown in Fig. 1, we consider a downlink MU-MIMO system enabled by MAs, where a BS equipped with $N$ transmit MAs serves $M$ single-antenna users. 
The MAs are connected to RF chains via flexible cables, which allows them to move freely within a designated local region $\mathcal{A}$ [16], [17]. Without loss of generality, we assume $\mathcal{A}$ a 2D square moving region of size $A\times A$. The position of the $n$-th transmit MA at the BS is represented by $\boldsymbol{t}_{n}=[x_n,y_n]^T$, and we define the set of positions as $\boldsymbol{t}\triangleq \left\{ \boldsymbol{t}_n \right\}$. The reference point of the region $\mathcal{A}$ is represented by $\boldsymbol{o}=[0,0]^T$.
The size of MAs' movement region is significantly smaller than the signal propagation distance, ensuring that the far-field condition holds between the BS and the users. In this configuration, altering the positions of the MAs affects only the phase of the complex channel coefficients for each channel path component without influencing the angle of departure (AoD), angle of arrival (AoA), or the amplitude of the channel gain.

The equivalent baseband channel from the BS to $m$-th user is denoted as ${{\boldsymbol{h}}_m}({\boldsymbol{t}}) \in {{\mathbb{C}}^{N \times 1}}$, characterized by the general Rician fading model
\begin{align}
\boldsymbol{h}_m\left( \boldsymbol{t} \right) =\sqrt{\frac{\kappa _m\beta _m}{\kappa _m+1}}\boldsymbol{\bar{h}}_m\left( \boldsymbol{t} \right) +\sqrt{\frac{\beta _m}{\kappa _m+1}}\boldsymbol{\tilde{h}}_m(\boldsymbol{t}) ,
\end{align}
where $\kappa_m \geq 0$ is the Rician factor, $\beta _m$ represents the large-scale fading coefficient, $\bar{\boldsymbol{h}} _m (\boldsymbol{t}) \in \mathbb{C}^{N \times 1}$ denotes the deterministic LoS component, and $\tilde{\boldsymbol{h}}_m (\boldsymbol{t}) \in \mathbb{C}^{N \times 1}$ represents the random non-line-of-sight (NLoS) component. Specifically, the NLoS component $\tilde{\boldsymbol{h}}_m (\boldsymbol{t})$ should be modeled as a spatially correlated Gaussian process with a covariance matrix $\tilde{\boldsymbol{h}}_m (\boldsymbol{t})=\boldsymbol{S}(\boldsymbol{t})\boldsymbol{u}_m$ with the Gaussian random variables $\boldsymbol{u}_m \sim \mathcal{C}\mathcal{N}\left( 0,\boldsymbol{I} \right)$,
and the spatial correlation matrix $\boldsymbol{S}\left( \boldsymbol{t} \right)$, which captures spatial dependencies due to the scattering environment. A widely adopted model for $\boldsymbol{S}(\boldsymbol{t}$ is the Bessel function model [12],[13], where the correlation between two antennas is expressed as $\left[ \boldsymbol{S}(\boldsymbol{t})\right]_{n,n'}=J_0\left( \frac{2\pi\lVert \boldsymbol{t}_{n}-\boldsymbol{t}_{n'} \rVert}{\lambda} \right) , \forall n,n'\in \mathcal{N}$, where $J_0(\cdot)$ is the zero-order Bessel function of the first kind, $\lVert \boldsymbol{t}_{n}-\boldsymbol{t}_{n'} \rVert$ is the distance between $n$-th and $n'$-th elements on the antenna array. For analytical tractability in our study, we simplify the NLoS component to i.i.d. complex Gaussian variables with zero mean and unit variance, approximating $\boldsymbol{S}(\boldsymbol{t})$ as an identity matrix. This simplification is supported by the imposed minimum antenna spacing constraint, which reduces spatial correlation, with a reasonable approximation error. We extend this analysis in Sections V and VI to evaluate the impact of spatial correlation, and in the following analysis and design, we simplify the NLoS modeling by assuming that the entries of $\tilde{\boldsymbol{h}}_m$ are independent and identically distributed (i.i.d.) circularly symmetric complex Gaussian random variables with zero mean and unit variance. 
To describe the LoS component, let $\theta _m\in \left[ -\frac{\pi}{2},\frac{\pi}{2} \right]$ and $\phi _m\in \left[ -\frac{\pi}{2},\frac{\pi}{2} \right]$ represent the elevation and azimuth AoDs for user $m$, respectively.
The propagation distance difference of the LoS path between the MA position $\boldsymbol{t}$ and the reference position $\boldsymbol{o}$ is given by
$\boldsymbol{t}_{n}^{T}\left[ \cos \theta _{m}^{}\sin \phi _{m}^{},\sin \theta _{m}^{} \right] ^T\triangleq\boldsymbol{t}_{n}^{T}\boldsymbol{a}_m$, 
and the phase difference of the LoS path is $\frac{2\pi}{\lambda}\boldsymbol{t}_{n}^{T}\boldsymbol{a}_m$, with $\lambda$ denoting the carrier wavelength.
Thus, the channel vector of the LoS component between the BS and $m$-th user can be expressed as
\begin{align}
\boldsymbol{\bar{h}}_m\left( \boldsymbol{t} \right) =\big[ e^{j\frac{2\pi}{\lambda}\boldsymbol{t}_{1}^{T}\boldsymbol{a}_m},\dots ,e^{j\frac{2\pi}{\lambda}\boldsymbol{t}_{N}^{T}\boldsymbol{a}_m} \big] ^T  .
\end{align}

\setlength{\abovecaptionskip}{0pt}
\begin{figure}
\centering
\includegraphics[width=3.4in]{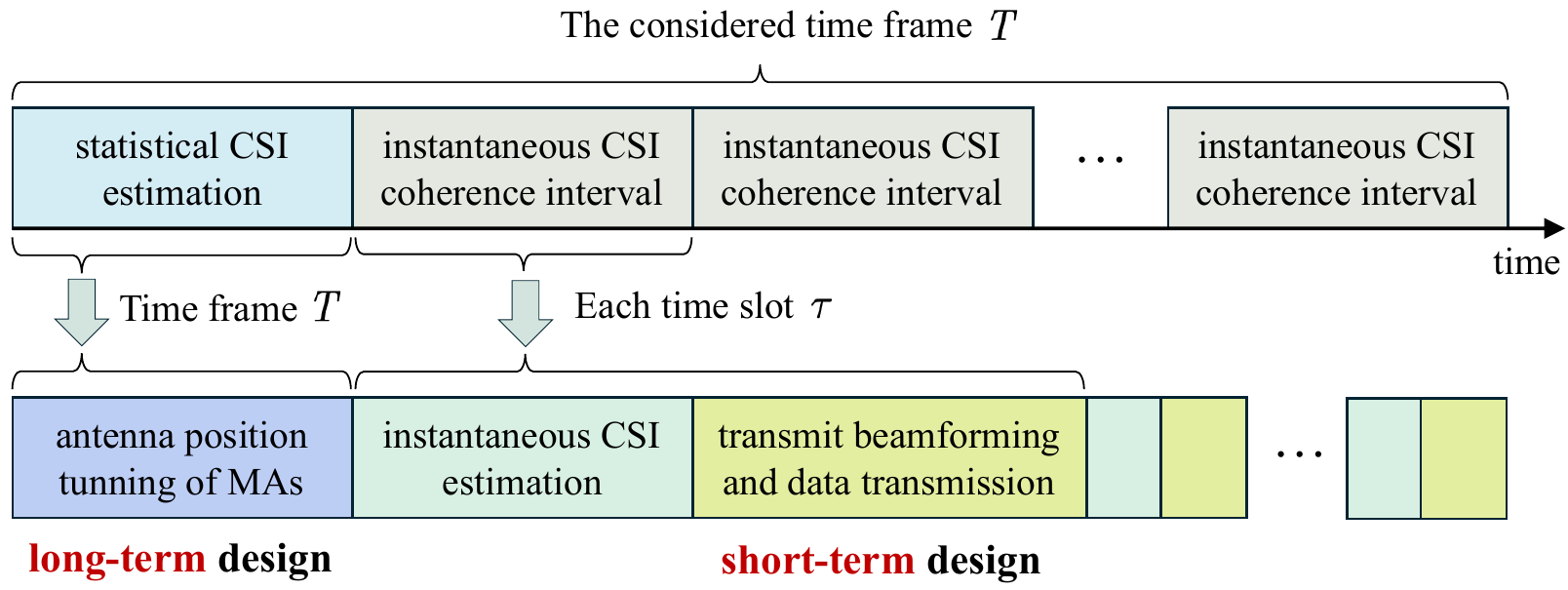}
\caption{Illustration of the proposed two-timescale framework.} \label{fig2}
\vspace{-18pt}
\end{figure}

\vspace{-3pt}
Let $\boldsymbol{W}=\left[ \boldsymbol{w}_1,\boldsymbol{w}_2,\cdots,\boldsymbol{w}_M \right] \in \mathbb{C} ^{N\times M}$ denote the linear precoding matrix at the BS, where each user is assigned a dedicated beamforming vector. The baseband transmitted signal is given by $\boldsymbol{Ws}$, where $\boldsymbol{s}=\left[ s_1,\cdots,s_M \right] ^T$ is the data vector, with each element $s_m$ being an independent variable with zero mean and normalized power.
The received signal at user $m$ can be expressed as
\begin{align}
u_m=\boldsymbol{h}_{m} ( \boldsymbol{t} ) ^{H}\boldsymbol{w}_m s_m+\sum_{j=1,j\ne m}^M{\boldsymbol{h}_m ( \boldsymbol{t}  ) ^{H}\boldsymbol{w}_j} s_j+z_m, 
\end{align}
where $z_m\sim \mathcal{C} \mathcal{N} \left( 0,\sigma_m^2 \right)$ represents the additive white Gaussian noise (AWGN) at the receiver of user $m$.
Accordingly, the SINR at user $m$ is given by
\begin{align}
\gamma_m=\frac{\left| \boldsymbol{h}_{m}( \boldsymbol{t} ) ^{H}\boldsymbol{w}_m \right|^2}{\sum_{j=1,j\ne m}^M{\left| \boldsymbol{h} _{m}( \boldsymbol{t})^{H}\boldsymbol{w}_j \right|^2}+\sigma _{m}^{2}}.
\end{align}
The sum achievable rate for all users is then expressed as 
\vspace{-3pt}
\begin{equation}
R= \sum_{m=1}^M{R_m}\triangleq\sum_{m=1}^M{\log _2\left( 1+\gamma_m \right)}.
\end{equation}
The sum rate $R$ of the MA-enabled downlink MU-MIMO system in (5) depends on the positions of the transmit MAs $\boldsymbol{t}$, which affect the BS-user channel $\boldsymbol{h}_{m}(\boldsymbol{t})$ as well as the corresponding transmit beamforming matrix $\boldsymbol{W}$.

\vspace{-10pt}
\subsection{Two-Timescale Transmission Scheme}
\vspace{-4pt}
In the Rician fading channel $\boldsymbol{h}_m\left( \boldsymbol{t} \right) $, the NLoS component $\boldsymbol{\tilde{h}}_m(\boldsymbol{t})$ is random and varies with different antenna positions at the BS. Estimating all $\{ \boldsymbol{\tilde{h}}_m\left( \boldsymbol{t} \right) \}_{m = 1}^M$ for arbitrary $\boldsymbol{t}$ would require the MA to traverse the entire feasible antenna region, making instantaneous CSI acquisition across the entire region impractical. 
In contrast, the LoS component in Rician fading channels remains relatively stable, primarily determined by spatial angles and position information. Additionally, large-scale path-loss coefficients and the Rician factor are static over a certain period, making them well-suited for statistical characterization. These properties allow for the optimization of antenna positions using statistical CSI, significantly reducing the overhead associated with real-time CSI estimation.

Building on this insight, we propose a hierarchical two-timescale transmission scheme, as shown in Fig. 2. In the first stage, the antenna positions at the BS are optimized based on statistical CSI, leveraging the stability of the LoS component to enhance long-term system performance\footnote{Incorporating ray tracing–based predictive CSI earlier, more precise environmental data refine position-based CSI, reduces CSI estimation overhead, and strengthens our two-timescale MA framework without altering its design.}. 
Once the optimal antenna positions are determined and fixed, the system transitions to the small timescale phase, where instantaneous CSI is acquired using conventional channel estimation techniques in multiple-antenna systems. This instantaneous CSI is then used to design the downlink beamforming, ensuring that the dynamic NLoS channel fluctuations are addressed in real time. 

Compared to schemes requiring continuous MA repositioning based on instantaneous CSI, our two-timescale design effectively reduces the update frequency of MAs' positions\footnote{This design is particularly beneficial in environments with rapid NLoS fluctuations but stable LoS information, while in high-mobility scenarios, the MA positions may still need to be updated more frequently based on changing statistical CSI to maintain performance. However, the two-timescale design continues to offer practical advantages by significantly reducing instantaneous CSI overhead.}., lowering channel estimation overhead and movement energy consumption\footnote{For a quantitative energy trade-off analysis, we model each MA’s repositioning energy as $E_n=P_x\frac{|\Delta x_n |}{v_x}+P_y\frac{| \Delta y_n|}{v_y}$ [43], where $P_x=P_y = 8\,\mathrm{W}$ (typical stepper motor power [44]) and $v_x=v_y\approx 0.94\,\mathrm{mm/ms}$ (based on a 10 mm lead screw [43]). In an MA array of $N=6$ with an average movement of $| \Delta x_n|=| \Delta y_n|=\lambda/2 = 25\,\mathrm{mm}$ at 12 GHz, repositioning consumes $E_{\mathrm{MA}} \approx 1.28\,\mathrm{J}$ per cycle, whereas data transmission at $P_{\mathrm{TR}} = 0.1\,\mathrm{W}$ over $300\,\mathrm{ms}$ uses only $E_{\mathrm{TR}} = 0.03\,\mathrm{J}$, making $E_{\mathrm{MA}}$ 42 times higher if executed every frame. By reducing the repositioning frequency (e.g., from every 300 ms to every 3 s) under stable statistical CSI, a two-timescale framework cuts $E_{\mathrm{MA}}$ by a factor of 10 while preserving performance.}. By integrating the NLoS component into the beamforming process after antenna positions are fixed, our method preserves the performance gains from this random component, combining the advantages of both statistical and instantaneous CSI. This two-timescale approach thus strikes an effective balance between practicality and performance.

\vspace{-12pt}
\subsection{Problem Formulation}
\vspace{-3pt}
Under the proposed two-timescale transmission framework, the objective is to maximize the ergodic sum rate for all users by jointly optimizing the antenna positions $\{{\boldsymbol{t}_n}\}$ in the large timescale and transmit beamforming vectors $\{{\boldsymbol{w}_m}\}$ in the small timescale at the BS. This leads to the formulation of a general optimization problem as follows\footnote{The two-timescale framework can be readily extended to multi-cell networks, where inter-cell interference becomes a key consideration, and remains sufficiently general to be adapted for various advanced wireless scenarios, e.g., machine-type communication, wireless sensing, and indoor localization.}
\vspace{-3pt}
\begin{subequations}
\begin{align}
\left( \text{P}1 \right) :\,\,\underset{\left\{ \boldsymbol{t}_n \right\}}{\max}&\,\,\mathbb{E}\bigg[ \underset{\left\{ \boldsymbol{w}_m \right\}}{\max}\sum_{m=1}^M{\log _2\left( 1+\gamma_m \right)} \bigg] \,\,  
\\
\text{s}.\text{t}.& \,\, \sum_{m=1}^M{\lVert \boldsymbol{w}_m \rVert ^2\le P_{\text{tot}}},\,\,
\\
&\,\,  \lVert \boldsymbol{t}_n-\boldsymbol{t}_i \rVert \ge D_{\min},\forall n,i\in \mathcal{N},i\ne n,\,
\\
&\,\,\, \boldsymbol{t}_n\in \mathcal{C},\forall n\in \mathcal{N},
\end{align}
\end{subequations}
where ${\mathbb{E}}\left[  \cdot  \right]$ in (6a) represents the expectation over all possible channel realizations, $P_{\text{tot}} \ge 0$ in (6b) denotes the maximum transmit power of the BS, (6c) ensures a minimum separation distance between any two adjacent MAs to avoid coupling effects and mitigate spatial correlation, and (6d) specifies the feasible region for MAs' positions.

Solving (P1) presents significant challenges due to the following reasons: 1. the long-term antenna positions variables $\{{\boldsymbol{t}_n}\}$ and short-term transmit beamforming variables $\{{\boldsymbol{w}_m}\}$ are intertwined in the objective function; and 2. the ergodic sum rate lacks a closed-form expression, especially with an unfixed set of antenna positions ${\boldsymbol{t}_n}$. Generally, there exists no efficient method to solve this non-convex problem optimally. For the general multiuser case, a stochastic optimization framework similar to the one in [32] can be employed to approximate the solution using randomly generated channel samples. However, this method incurs high computational complexity depending on the scale of the channel samples.

Motivated by these observations, in the following two sections, we investigate the transmission designs for two classical beamforming schemes, namely MRT beamforming and ZF beamforming, in the context of (P1), respectively. For each scheme, we first derive the achievable rate expression and then propose an efficient algorithm for optimizing the antenna positions, followed by computational complexity analysis.

\vspace{-9pt}
\section{Achievable Rate Derivation and Antenna Position Optimization under MRT Beamforming}
\vspace{-3pt}
In this section, within the proposed two-timescale scheme, we derive a closed-form expression for the approximate ergodic rate, leveraging statistical CSI and the structure of small-timescale MRT beamforming under a fixed power allocation strategy. Then, using these expressions and resorting to AO and SCA techniques, we develop an iterative algorithm for antenna position optimization. The analysis and design are extended to ZF beamforming in Section IV.

\vspace{-9pt}
\subsection{Ergodic Rate Analysis with MRT beamforming}
\vspace{-3pt}
The MRT precoding vector, assuming perfect knowledge of the instantaneous CSI, is expressed as
\begin{align}
\boldsymbol{w}_{\text{MRT},m}=\sqrt{p_m}\boldsymbol{h}_{m}\left( \boldsymbol{t} \right) , 
\end{align}
where $p_m$ represents the power allocation parameter, determined to satisfy the total beam power constraint
\begin{align}
\sum_{m=1}^M{\lVert \boldsymbol{w}_{\text{MRT},m}\left( \boldsymbol{t} \right) \rVert ^2}
\triangleq\sum_{m=1}^M{p_m\lVert \boldsymbol{h}_{m}^{}\left( \boldsymbol{t} \right) \rVert ^2}
\le P_{\text{tot}}.
\end{align}
The SINR for MRT beamforming is then given by 
\begin{align}
\gamma _{\text{MRT},m}=\frac{p_m\lVert \boldsymbol{h}_{m}^{}\left( \boldsymbol{t} \right) \rVert ^4}{\sum_{j=1,j\ne m}^M{p_j}\left| \boldsymbol{h}_{j}^{H}\left( \boldsymbol{t} \right) \boldsymbol{h}_m\left( \boldsymbol{t} \right) \right|^2+\sigma _{m}^{2}}. 
\end{align}
Note that MRT beamforming aligns conjugate vectors with the corresponding BS-user channels, followed by power allocation\footnote{In the two-timescale MA transmission scheme, power allocation can also be optimized based on statistical CSI in the large timescale. However, due to limited gains relative to the added complexity, the joint power allocation and antenna position optimization are left for our future work.}. To underscore the potential of MA and its position optimization, we simplify the power allocation process to a fixed scheme by setting it proportional to the channel gain between the BS and each user, as follows
\begin{align}
p_m=p\triangleq\frac{P_{\text{tot}}}{\sum_{m=1}^M{\lVert \boldsymbol{h}_{m}^{}\left( \boldsymbol{t} \right) \rVert ^2}}.
\end{align}

With MRT beamforming, the antenna position optimization problem for ergodic sum rate maximization can be recast as
\begin{align}
\left( \text{P}2 \right) :\underset{\left\{ \boldsymbol{t}_n \right\}}{\max}&\,\,\mathbb{E}\left[ \log _2\left( 1+\gamma _{\text{MRT},m} \right) \right] 
\\
\text{s}.\text{t}.&\,\, (6\text{c}), (6\text{d}). \nonumber 
\end{align}
Based on the approximation in [33, Lemma 1], the ergodic rate for user $m$ under statistical CSI is expressed as
\begin{align}
\mathbb{E}&\left\{ \log _2\left( 1+\gamma _{\text{MRT},m} \right) \right\} \nonumber
\\
&\approx \log _2\bigg(\! 1 \!+\!\frac{\mathbb{E}\left\{ \lVert \boldsymbol{h}_{m}^{}\left( \boldsymbol{t} \right) \rVert ^4 \right\}}{\mathbb{E}\big\{ \sum_{j=1,j\ne m}^M{| \boldsymbol{h}_{j}^{H}\left( \boldsymbol{t} \right) \boldsymbol{h}_m\left( \boldsymbol{t} \right) |^2} \big\} \!+\!\mathbb{E}\{ \frac{\sigma _{m}^{2}}{p} \}} \!\bigg)
\nonumber
\\
&\triangleq R_{\text{MRT},m}^{app}.
\end{align}
This approximation lies between the upper and lower bounds of $\mathbb{E}\left\{ \log _2\left( 1+\gamma _{\text{MRT},m} \right) \right\}$. After calculating the expectation terms, we arrive at (13) provided at the top of the next page.
\begin{figure*}
\begin{align}
& \quad\quad\quad\quad\quad R_{\text{MRT},m}^{app}=\log _2\Bigg( 1+\frac{\beta _{m}^{2}\left( \frac{2N\kappa _m+N}{\left( \kappa _m+1 \right) ^2}+N^2 \right)}{\sum_{j=1,j\ne m}^M{\beta_m\beta_j\frac{\kappa_m\kappa _j\left| \boldsymbol{\bar{h}}_j\left( \boldsymbol{t} \right) \boldsymbol{\bar{h}}_{m}^{H}\left( \boldsymbol{t} \right) \right|^2+N\left( \kappa _m+\kappa _j+1 \right)}{\left( \kappa _m+1 \right) \left( \kappa _j+1 \right)}}+\frac{\sigma _{m}^{2}}{P}N\sum_{m=1}^M{\beta _{m}^{}}} \Bigg).
\end{align}
\vspace{-24pt}
\end{figure*}

\textit{Remark 1:} From (13), we can draw three key insights. 
First, under MRT beamforming in the two-timescale design,  when $M=1$ there is no ergodic rate gain from antenna position tuning. This is because the gain originates from interference mitigation or channel decorrelation between $\boldsymbol{\bar{h}}_j\left( \boldsymbol{t} \right)$ and $ \boldsymbol{\bar{h}}_{m}^{H}\left( \boldsymbol{t} \right)$, $\forall j \ne m$, which is achieved by optimizing $\boldsymbol{t}$ to tune the phases of the LoS components and reduce $\left| \boldsymbol{\bar{h}}_j\left( \boldsymbol{t} \right) \boldsymbol{\bar{h}}_{m}^{H}\left( \boldsymbol{t} \right) \right|^2$.
Second, with $\kappa_m=0$ in the Rician fading channel, the absence of a deterministic LoS component will render MA ineffective in reshaping the channel since the summation term in the denominator of (13) vanishes, and thus $R_{\text{MRT},m}^{app}$ no longer related to $\boldsymbol{t}$.
Third, as the number of antennas $N$ becomes asymptotically large, i.e., $N\rightarrow \infty$, the performance gain of MA diminishes due to the channel hardening, where FPAs already achieve asymptotically orthogonal channels for different users, i.e., $\big|\frac{1}{N}\boldsymbol{\bar{h}}_j\left( \boldsymbol{t}_{\text{FPA}} \right) \boldsymbol{\bar{h}}_{m}^{H}\left( \boldsymbol{t}_{\text{FPA}} \right) \big|^2\rightarrow0, \forall j,m$.

\vspace{-16pt}
\subsection{Antenna Position Optimization Based on AO Method}
\vspace{-5pt}
Although we obtain the closed-form expression (13), MAs' positions remain intricately coupled in the non-convex term $|\boldsymbol{\bar{h}}_j\left( \boldsymbol{t} \right) \boldsymbol{\bar{h}}_{m}^{H}\left( \boldsymbol{t} \right)|^2$ relying on the LoS components. The non-convex minimum distance constraint (6c) further complicates the problem (P2), making it challenging to solve directly.

To address this challenge, we propose an antenna-wise position optimization framework based on the AO method. In this framework, we first derive a tractable expression for the achievable rate concerning the $n$-th antenna while keeping the positions of the other antennas fixed. 
Focusing on the $n$-th antenna, we can express the critical term $\big| \boldsymbol{\bar{h}}_j\left( \boldsymbol{t} \right) \boldsymbol{\bar{h}}_{m}^{H}\left( \boldsymbol{t} \right) \big|^2=\big| \sum_{n=1}^N{e^{j\frac{2\pi}{\lambda}\boldsymbol{t}_{n}^{T}\left( \boldsymbol{a}_m-\boldsymbol{a}_j \right)}} \big|^2$ as
\begin{align}
&\!\big| \boldsymbol{\bar{h}}_j\left( \boldsymbol{t}_n \right) \boldsymbol{\bar{h}}_{m}^{H}\left( \boldsymbol{t}_n \right) \big|^2 \nonumber
\\
&\!=2\text{Re} \{ e^{j\frac{2\pi}{\lambda}\boldsymbol{t}_{n}^{T}\boldsymbol{a}_{m,j}}\tau_{n,m,j}^{*} \} +\left| \tau_{n,m,j} \right|^2+1 \nonumber
\\
&\!=\!2\left| \tau_{n,m,j}^{} \right|\cos \big( \frac{2\pi}{\lambda}\boldsymbol{t}_{n}^{T}\boldsymbol{a}_{m,j}\!-\!\angle \tau_{n,m,j}^{} \big)\! +\!\left| \tau_{n,m,j} \right|^2\!+\!1, \!\!\!\!\!
\end{align}
with $\tau_{n,m,j}=\sum_{i=1,i\ne n}^N{e^{j\frac{2\pi}{\lambda}\boldsymbol{t}_{i}^{T}\left( \boldsymbol{a}_m-\boldsymbol{a}_j \right)}}$. Based on (13) and (14), the ergodic rate of the $m$-th user is given by (15) at the top of this page,
\begin{figure*}
\begin{align}
&\quad \quad \quad \quad \quad \quad R_{\text{MRT},m}^{app}(\boldsymbol{t}_n)=\log _2\Big( 1+ \frac{c_{2,m}}{2\sum_{j=1,j\ne m}^M{c_{1,m,j}\left| \tau_{n,m,j}^{} \right|\cos \left( \frac{2\pi}{\lambda}\boldsymbol{t}_{n}^{T}\boldsymbol{a}_{m,j}\!-\!\angle \tau_{n,m,j}^{} \right)}\!+\!c_{3,m}}\! \Big), 
\\
&\overline{\ \ \ \ \ \ \ \ \ \ \ \ \ \ \ \ \ \ \ \ \ \ \ \ \ \ \ \ \ \ \ \ \ \ \ \ \ \ \ \ \ \ \ \ \ \ \ \ \ \ \ \ \ \ \ \ \ \ \ \ \ \ \ \ \ \ \ \ \ \ \ \ \ \ \ \ \ \ \ \ \ \ \ \ \ \ \ \ \ \ \ \ \ \ \ \ \ \ \ \ \ \ \ \ \ \ \ \ \ \ \ \ \ \ \ \ \ \ \ \ \ \ \ \ \ \ \ \ \ \ \ \ \ \ \ \ \ \ \ \ \ \ \ \ } \nonumber
\end{align}
\vspace{-33pt}
\end{figure*}
where parameters $c_{1,m,j},c_{2,m}$, and $c_{3,m}$ are respectively defined as
\begin{subequations}
\begin{align}
&c_{1,m,j}=\frac{\beta _{j}^{}\kappa _m\kappa _j}{\left( \kappa _m+1 \right) \left( \kappa _j+1 \right)},
\\
&c_{2,m}=\beta _{m}^{}\Big( \frac{2N\kappa _m+N}{\left( \kappa _m+1 \right) ^2}+N^2 \Big) ,
\\
&c_{3,m}=\sum_{j=1,j\ne m}^M{c_{1,m,j}\left( \left| \tau_{n,m,j} \right|^2+1 \right)}+\frac{\sigma _{m}^{2}N}{\beta_m P_{\text{tot}}}\sum_{j=1}^M{\beta_j} \nonumber
\\
&\quad \quad \quad +\sum_{j=1,j\ne m}^M{\frac{N\beta _{j}^{}\left( \kappa _m+\kappa _j+1 \right)}{\left( \kappa _m+1 \right) \left( \kappa _j+1 \right)}}.
\end{align}
\end{subequations}

The antenna position optimization problem for $\boldsymbol{t}_n$ with MRT beamforming can now be reformulated as
\begin{align}
\left( \text{P}2.n \right) : \,\, \underset{\boldsymbol{t}_n}{\max}&\sum_{m=1}^M{R_{\text{MRT},m}^{app}\left( \boldsymbol{t}_n \right)}
\\
 \text{s}.\text{t}. &\,\, ( 6\text{c}), (6\text{d}). \nonumber
\end{align}
Although the objective function has been simplified, the term $2\sum_{j=1,j\ne m}^M{c_{1,m,j}\left| \tau_{n,m,j}^{} \right|\cos \left( \frac{2\pi}{\lambda}\boldsymbol{t}_{n}^{T}\boldsymbol{a}_{m,j}\!-\!\angle \tau_{n,m,j}^{} \right)}$ in (15), along with its inverse, remains neither concave nor convex with respect to $\boldsymbol{t}_{n}$. This non-convexity poses a challenge, preventing the construction of a global lower bound for the objective function using the first-order Taylor expansion.

\vspace{-16pt}
\subsection{Constructing a Surrogate Function via SCA Technique}
\vspace{-6pt}
To address the non-convexity of the objective function in (17), we propose constructing a concave lower-bound surrogate function for $\boldsymbol{t}_{n}$ by sequentially applying first-order and second-order Taylor expansions. 

First, we define the function $b_m(\boldsymbol{t}_n) $, representing the non-convex term in the denominator of (15)
\vspace{-3pt}
\begin{align}
&\!\!\!b_m(\boldsymbol{t}_n) = \nonumber
\\
&\quad 2\sum_{j=1,j\ne m}^M{c_{1,m,j}  \left| \tau_{n,m,j}^{} \right|\cos \Big( \frac{2\pi}{\lambda}\boldsymbol{t}_{n}^{T}\boldsymbol{a}_{m,j}  \!-  \!\angle \tau_{n,m,j}^{} \Big)}.\!\!
\end{align}
The expression $R_{\text{MRT},m}^{app}(\boldsymbol{t}_n) \!=\!\log _2\big( 1\!+\!\frac{c_{2,m}}{b_m\left( \boldsymbol{t}_{n}^{} \right) +c_{3,m}} \big) $ is convex with respect to $b_m(\boldsymbol{t}_n)$. Using the first-order Taylor expansion [34], we derive the following concave lower bound
\begin{align}
&\log _2\Big( 1+\frac{c_{2,m}}{b_m\left( \boldsymbol{t}_{n}^{} \right) +c_{3,m}}\Big) \ge \log _2\Big( 1+\frac{c_{2,m}}{b_m\left( \boldsymbol{t}_{n}^{\ell} \right) +c_{3,m}} \Big) \nonumber
\\
&\qquad\qquad -\frac{c_{2,m}\log _2e}{\left( b_m\left( \boldsymbol{t}_{n}^{\ell} \right) +c_{3,m} \right) \left( b_m\left( \boldsymbol{t}_{n}^{\ell} \right) +c_{2,m}+c_{3,m} \right)} \nonumber
\\
&\qquad\qquad\quad \times
\left( b_m\left( \boldsymbol{t}_{n}^{} \right) -b_m\left( \boldsymbol{t}_{n}^{\ell} \right) \right), 
\end{align}
where $\boldsymbol{t}_{n}^{\ell}$ represents the local point at the $\ell$-th iteration in the SCA process. Nevertheless, $b_m( \boldsymbol{t}_{n}^{\ell})$ is still neither concave nor convex with respect to $\boldsymbol{t}_{n}$, requiring further refinement to achieve a more tractable optimization framework.

To overcome this challenge, we apply the second-order Taylor expansion (also known as the Descent Lemma [35, Lemma 12]) to construct a convex upper-bound surrogate function for $b_m( \boldsymbol{t}_{n}^{\ell})$. Specifically, we introduce a positive scalar $\psi_{m,n}$ such that $\psi _{m,n}\boldsymbol{I}\succeq \nabla ^2b_m\left( \boldsymbol{t}_{n}^{} \right)$. This ensures the following inequality
\begin{align}
b_m\left( \boldsymbol{t}_{n}^{} \right) &\le b_m\left( \boldsymbol{t}_{n}^{\ell} \right) +\nabla b_m\left( \boldsymbol{t}_{n}^{\ell} \right) ^T\left( \boldsymbol{t}_{n}^{}-\boldsymbol{t}_{n}^{\ell} \right) \nonumber
\\
&\quad+\frac{\psi _{m,n}}{2}\left( \boldsymbol{t}_{n}^{}-\boldsymbol{t}_{n}^{\ell} \right) ^T\left( \boldsymbol{t}_{n}^{}-\boldsymbol{t}_{n}^{\ell} \right) \nonumber
\\
&\triangleq g_{m}^{\text{ub},\ell}\left( \boldsymbol{t}_{n}^{} \right),  
\end{align}
where the gradient $\nabla b_m\left( \boldsymbol{t}_{n}^{\ell} \right) $ is computed as
\begin{align}
\nabla b_m\left( \boldsymbol{t}_{n}^{\ell} \right) \,\,=\left[ \frac{\partial b_m\left( \boldsymbol{t}_{n}^{} \right)}{\partial x_n}\Big|_{\boldsymbol{t}_{n}^{}=\boldsymbol{t}_{n}^{\ell}}^{},\frac{\partial b_m\left( \boldsymbol{t}_{n}^{} \right)}{\partial y_n}\Big|_{\boldsymbol{t}_{n}^{}=\boldsymbol{t}_{n}^{\ell}}^{} \right],
\end{align}
with the detailed expressions of $\frac{\partial b_m\left( \boldsymbol{t}_{n}^{} \right)}{\partial x_n}\mid_{\boldsymbol{t}_{n}^{}=\boldsymbol{t}_{n}^{\ell}}^{}$ and $\frac{\partial b_m\left( \boldsymbol{t}_{n}^{} \right)}{\partial y_n}\mid_{\boldsymbol{t}_{n}^{}=\boldsymbol{t}_{n}^{\ell}}^{} $ provided in (22) at the top of this page.
\begin{figure*}
\begin{subequations}
\begin{align}
&\frac{\partial b_m\left( \boldsymbol{t}_{n}^{} \right)}{\partial x_n}\Big|_{\boldsymbol{t}_{n}^{}=\boldsymbol{t}_{n}^{\ell}}^{}=-\frac{4\pi}{\lambda}\sum_{j=1,j\ne m}^M{c_{1,m,j}\left| \tau_{n,m,j}^{} \right|\left( \cos \theta _{m}^{}\sin \phi _{m}^{}-\cos \theta _{j}^{}\sin \phi _{j}^{} \right) \sin \left( \frac{2\pi}{\lambda}\boldsymbol{t}_{n}^{T}\left( \boldsymbol{a}_m-\boldsymbol{a}_j \right) -\angle \tau_{n,m,j}^{} \right)},
\\
&\frac{\partial b_m\left( \boldsymbol{t}_{n}^{} \right)}{\partial y_n}\Big|_{\boldsymbol{t}_{n}^{}=\boldsymbol{t}_{n}^{\ell}}^{}=-\frac{4\pi}{\lambda}\sum_{j=1,j\ne m}^M{c_{1,m,j}\left| \tau_{n,m,j}^{} \right|\left( \sin \theta _{m}^{}-\sin \theta _{j}^{} \right) \sin \left( \frac{2\pi}{\lambda}\boldsymbol{t}_{n}^{T}\left( \boldsymbol{a}_m-\boldsymbol{a}_j \right) -\angle \tau_{n,m,j}^{} \right)}.
\\
&\overline{\ \ \ \ \ \ \ \ \ \ \ \ \ \ \ \ \ \ \ \ \ \ \ \ \ \ \ \ \ \ \ \ \ \ \ \ \ \ \ \ \ \ \ \ \ \ \ \ \ \ \ \ \ \ \ \ \ \ \ \ \ \ \ \ \ \ \ \ \ \ \ \ \ \ \ \ \ \ \ \ \ \ \ \ \ \ \ \ \ \ \ \ \ \ \ \ \ \ \ \ \ \ \ \ \ \ \ \ \ \ \ \ \ \ \ \ \ \ \ \ \ \ \ \ \ \ \ \ \ \ \ \ \ \ \ \ \ \ \ \ \ \ \ \ } \nonumber
\end{align}
\end{subequations}
\vspace{-36pt}
\end{figure*}
Next, we determine the scalar $\psi _{m,n}$ such that it meets the constraint $\psi _{m,n}\boldsymbol{I}\succeq \nabla ^2b_m\left( \boldsymbol{t}_{n}^{} \right)$, ensuring a suitable upper-bound approximation. To start with, we calculate the Hessian matrix $\nabla ^2b_m\left( \boldsymbol{t}_{n}^{} \right)$, which is expressed as
\begin{align}
\nabla ^2b_m\left( \boldsymbol{t}_{n}^{} \right) \,\,=\left[ \begin{matrix}
	\frac{\partial b_m\left( \boldsymbol{t}_{n}^{} \right)}{\partial x_n\partial x_n}&		\frac{\partial b_m\left( \boldsymbol{t}_{n}^{} \right)}{\partial x_n\partial y_n}\\
	\frac{\partial b_m\left( \boldsymbol{t}_{n}^{} \right)}{\partial y_n\partial x_n}&		\frac{\partial b_m\left( \boldsymbol{t}_{n}^{} \right)}{\partial y_n\partial y_n}\\
\end{matrix} \right], 
\end{align}
with the detailed entries of this matrix provided in (24) at the top of this page.
\begin{figure*}
\begin{subequations}
\begin{align}
&\frac{\partial b_m\left( \boldsymbol{t}_{n}^{} \right)}{\partial x_n\partial x_n}=-\frac{8\pi ^2}{\lambda ^2}\sum_{j=1,j\ne m}^M{c_{1,m,j}\left| \tau_{n,m,j}^{} \right|\left( \cos \theta _{m}^{}\sin \phi _{m}^{}-\cos \theta _{j}^{}\sin \phi _{j}^{} \right) ^2\cos \left( \frac{2\pi}{\lambda}\boldsymbol{t}_{n}^{T}\left( \boldsymbol{a}_m-\boldsymbol{a}_j \right) -\angle \tau_{n,m,j}^{} \right)},
\\
&\frac{\partial b_m\left( \boldsymbol{t}_{n}^{} \right)}{\partial x_n\partial y_n}=\frac{\partial b_m\left( \boldsymbol{t}_{n}^{} \right)}{\partial y_n\partial x_n} \nonumber
\\
&=-\frac{8\pi ^2}{\lambda ^2}\sum_{j=1,j\ne m}^M{c_{1,m,j}\left| \tau_{n,m,j}^{} \right|\left( \cos \theta _{m}^{}\sin \phi _{m}^{}-\cos \theta _{j}^{}\sin \phi _{j}^{} \right) \left( \sin \theta _{m}^{}-\sin \theta _{j}^{} \right) \cos \left( \frac{2\pi}{\lambda}\boldsymbol{t}_{n}^{T}\left( \boldsymbol{a}_m-\boldsymbol{a}_j \right) -\angle \tau_{n,m,j}^{} \right)},
\\
&\frac{\partial b_m\left( \boldsymbol{t}_{n}^{} \right)}{\partial y_n\partial y_n}=-\frac{8\pi ^2}{\lambda ^2}\sum_{j=1,j\ne m}^M{c_{1,m,j}\left| \tau_{n,m,j}^{} \right|\left( \sin \theta _{m}^{}-\sin \theta _{j}^{} \right) ^2\cos \left( \frac{2\pi}{\lambda}\boldsymbol{t}_{n}^{T}\left( \boldsymbol{a}_m-\boldsymbol{a}_j \right) -\angle \tau_{n,m,j}^{} \right)}.
\\
&\overline{\ \ \ \ \ \ \ \ \ \ \ \ \ \ \ \ \ \ \ \ \ \ \ \ \ \ \ \ \ \ \ \ \ \ \ \ \ \ \ \ \ \ \ \ \ \ \ \ \ \ \ \ \ \ \ \ \ \ \ \ \ \ \ \ \ \ \ \ \ \ \ \ \ \ \ \ \ \ \ \ \ \ \ \ \ \ \ \ \ \ \ \ \ \ \ \ \ \ \ \ \ \ \ \ \ \ \ \ \ \ \ \ \ \ \ \ \ \ \ \ \ \ \ \ \ \ \ \ \ \ \ \ \ \ \ \ \ \ \ \ \ \ \ \ } \nonumber
\end{align}
\end{subequations}
\vspace{-36pt}
\end{figure*}
The basic scaling concept, as presented in [15] and [23], is expressed as
\begin{align}
\lVert \nabla ^2b_m\left( \boldsymbol{t}_{n}^{} \right) \rVert _2 &\le \lVert \nabla ^2b_m\left( \boldsymbol{t}_{n}^{} \right) \rVert _F \nonumber
\\
&\le \left\| \begin{matrix}
	\varPsi _{11}&		\varPsi _{12}\\
	\varPsi _{12}&		\varPsi _{22}\\
\end{matrix} \right\| _F \triangleq \bar{\psi}_{m,n}, 
\end{align}
where the matrix $\boldsymbol{\varPsi}\!\!=\!\!\left[\begin{matrix}
	\varPsi _{11}\!\!&	\!	\varPsi _{12}\\
	\varPsi _{12}\!\!&	\!	\varPsi _{22}\\
\end{matrix}\right]$ in (26) has entries defined as 
\vspace{-9pt}
\begin{subequations}
\begin{align}
\varPsi _{11}=&\sum_{j=1,j\ne m}^M{c_{1,m,j}\left| \tau_{n,m,j}^{} \right|\left( \cos \theta _{m}^{}\sin \phi _{m}^{}-\cos \theta _{j}^{}\sin \phi _{j}^{} \right) ^2},
\end{align}
\begin{align}
\varPsi _{12}=&\sum_{j=1,j\ne m}^M{c_{1,m,j}\left| \tau_{n,m,j}^{} \right| \big| ( \cos \theta _{m}^{}\sin \phi _{m}^{}-\cos \theta _{j}^{}\sin \phi _{j}^{} ) }
\nonumber
\\
&\times  ( \sin \theta _{m}^{}-\sin \theta _{j}^{} ) \big|,
\\
\varPsi _{22}=&\sum_{j=1,j\ne m}^M{c_{1,m,j}\left| \tau_{n,m,j}^{} \right|\left( \sin \theta _{m}^{}-\sin \theta _{j}^{} \right) ^2},
\end{align}
by simplifying $\cos \left( \frac{2\pi}{\lambda}\boldsymbol{t}_{n}^{T}\boldsymbol{a}_{m,j}-\angle \tau_{n,m,j}^{} \right) = 1,\forall j\ne m$, and taking absolute values for $\left( \cos \theta _{m}^{}\sin \phi _{m}^{}-\cos \theta _{j}^{}\sin \phi _{j}^{} \right)$ $\times \left( \sin \theta _{m}^{}-\sin \theta _{j}^{} \right) ,\forall j\ne m$, $\frac{\partial b_m\left( \boldsymbol{t}_{n}^{} \right)}{\partial x_n\partial x_n},\frac{\partial b_m\left( \boldsymbol{t}_{n}^{} \right)}{\partial x_n\partial y_n}$, and $\frac{\partial b_m\left( \boldsymbol{t}_{n}^{} \right)}{\partial y_n\partial y_n}$. 
In fact, the upper bound in (20), influenced by $\psi _{m,n}$, can be further tightened by selecting $\psi _{m,n}$ as follows
\end{subequations}
\begin{align}
\lVert \nabla ^2b_m\left( \boldsymbol{t}_{n}^{} \right) \rVert _2 &\le \left\| \begin{matrix}
	\big|\frac{\partial b_m\left( \boldsymbol{t}_{n}^{} \right)}{\partial x_n\partial x_n} \big|&		\big| \frac{\partial b_m\left( \boldsymbol{t}_{n}^{} \right)}{\partial x_n\partial y_n} \big|\\
	\big| \frac{\partial b_m\left( \boldsymbol{t}_{n}^{} \right)}{\partial y_n\partial x_n} \big|&		\big| \frac{\partial b_m\left( \boldsymbol{t}_{n}^{} \right)}{\partial y_n\partial y_n} \big|\\
\end{matrix} \right\| _2 \nonumber
\\
&\le \left\| \begin{matrix}
	\varPsi _{11}&		\varPsi _{12}\\
	\varPsi _{12}&		\varPsi _{22}\\
\end{matrix} \right\| _2 \triangleq \psi _{m,n},
\end{align}
still ensuring that $\psi _{m,n}\boldsymbol{I}\succeq \nabla ^2b_m\left( \boldsymbol{t}_{n}^{} \right)$.
Compared to $\bar{\psi}_{m,n}$ in (25), the refined bound satisfies 
\begin{align}
&\psi _{m,n}\triangleq \left\| \begin{matrix}
	\varPsi _{11}&		\varPsi _{12}\\
	\varPsi _{12}&		\varPsi _{22}\\
\end{matrix} \right\| _2\le \left\| \begin{matrix}
	\varPsi _{11}&		\varPsi _{12}\\
	\varPsi _{12}&		\varPsi _{22}\\
\end{matrix} \right\|_F\triangleq \bar{\psi}_{m,n}.
\end{align}
The refined bound $\psi _{m,n}$ provides a more accurate upper bound for the surrogate function used in the optimization process. 
Using the matrix 2-norm and eigenvalue formulas, $\psi _{m,n}$ can be readily computed as
\begin{subequations}
\begin{align}
\psi _{m,n}&=\frac{8\pi ^2}{\lambda ^2} \epsilon _{\max}\left( \boldsymbol{\varPsi} \right) 
\\
&=\frac{4\pi ^2}{\lambda ^2}\Big( \varPsi _{11}+\varPsi _{22}+\sqrt{\left( \varPsi _{11}-\varPsi _{22} \right) ^2-4\varPsi _{12}^{2}} \Big), 
\end{align}
\end{subequations}
where $\epsilon _{\max}(\cdot)$ denotes the maximum eigenvalue of a matrix. 

With the above transformation, we construct a convex upper-bound surrogate function for $b_m( \boldsymbol{t}_{n}^{\ell})$ and, consequently, derive a concave lower-bound surrogate function for (18). This approach, combined with the first-order approximation in (19), facilitates the design of an efficient antenna position optimization algorithm that iteratively refines the solution within the SCA framework.
The sole remaining hurdle in solving the problem (P2.n) lies in the non-convex minimum distance constraint (6d), which is equivalent to ensuring that $\lVert \boldsymbol{t}_n-\boldsymbol{t}_i \rVert ^2\ge  D_{\min}^2,\forall n,i\in \mathcal{N},i\ne n$. Given that $\lVert \boldsymbol{t}_n-\boldsymbol{t}_i \rVert ^2$ is convex with respect to $\boldsymbol{t}_n$, it is lower bounded by its first-order Taylor expansion as follows
\begin{align}
\lVert \boldsymbol{t}_n-\boldsymbol{t}_i \rVert ^2&\ge \lVert \boldsymbol{t}_{n}^{\ell}-\boldsymbol{t}_i \rVert ^2+2\left( \boldsymbol{t}_{n}^{\ell}-\boldsymbol{t}_i \right) ^T\left( \boldsymbol{t}_{n}^{}-\boldsymbol{t}_{n}^{\ell} \right)  \nonumber
\\
&\triangleq \mathcal{T}_{}^{\text{lb},\ell}\left( \boldsymbol{t}_n \right).    
\end{align}

\vspace{-12pt}
\subsection{Overall Algorithm and Computational Complexity Analysis}
\vspace{-3pt}
By substituting the objective function with the surrogate function that combines (18) and (19) and replacing $\lVert \boldsymbol{t}_n-\boldsymbol{t}_i \rVert ^2$ with $\mathcal{T}_{}^{\text{lb},\ell}\left( \boldsymbol{t}_n \right)$, the approximation optimization problem is
\begin{subequations}
\begin{align}
\text{(P3.n):}\,\,&\underset{\boldsymbol{t}_{n}}{\max}\,\,\sum_{m=1}^M{\log _2\left( 1+\frac{c_{2,m}}{b_m\left( \boldsymbol{t}_{n}^{\ell} \right) +c_{3,m}}\,\, \right)} \nonumber
\\
&\!\!\!\! -\frac{c_{2,m}\log _2e}{\left( b_m\left( \boldsymbol{t}_{n}^{\ell} \right) +c_{3,m} \right) \left( b_m\left( \boldsymbol{t}_{n}^{\ell} \right) +c_{3,m}+c_{2,m} \right)}\,\, \nonumber
\\
&\!\!\times \big( \nabla b_m\left( \boldsymbol{t}_{n}^{\ell} \right) ^T\left( \boldsymbol{t}_{n}^{}-\boldsymbol{t}_{n}^{\ell} \right) +\frac{\psi _{m,n}}{2}\lVert \boldsymbol{t}_{n}^{}-\boldsymbol{t}_{n}^{\ell} \rVert ^2 \big)  \!\!\!
\\
&\,\text{s.t.} \,\,\mathcal{T}^{\text{lb},\ell}( \boldsymbol{t}_n) \ge D_{\min}^{2},\forall n,i\in \mathcal{N},
\\
&\,\quad \,\,\,( 6\text{d}).  \nonumber
\end{align}
\end{subequations}
As a convex QCP, this problem can be solved efficiently using standard solvers such as CVX [36].
By solving (P3.n), we obtain a lower bound on the optimal value of (P2.n).

The overall AO-based algorithm is detailed in Algorithm 1, deriving a solution for the problem (P2) by iteratively solving (P3.n). The algorithm guarantees convergence to local optimal points since the objective value is non-decreasing in each iteration. The total computational complexity is $\mathcal{O}( L_1( N^{2.5}\log \frac{1}{\varepsilon}+NM^2+MN^2 ))$, derived as follows: for each antenna, a convex quadratic constrained program (QCP) is solved using an interior-point method, with complexity $\mathcal{O}(N^{1.5}\log \frac{1}{\varepsilon})$, where $\varepsilon$ denotes the prescribed accuracy. This accounts for the 2D optimization variable $\boldsymbol{t}_n=\left[ x_n,y_n \right] ^T$ with $\mathcal{O}(N)$ convexified distance constraints and linear region constraints. Per iteration, evaluate the surrogate objective costs $\mathcal{O}(M^2+MN)$, calculate the gradient costs $\mathcal{O}(M^2)$, and determine the Hessian bound costs $\mathcal{O}(M^2)$. The algorithm iterates over all $N$ antennas $L_1$ times until convergence, summing to $\mathcal{O}( L_1( N^{2.5}\log \frac{1}{\varepsilon}+NM^2+MN^2 ))$.

\setlength{\textfloatsep}{3pt}
\begin{algorithm}[t]
\caption{AO-based optimization algorithm for (P2).}\label{alg:alg1}
\begin{algorithmic}[1]
\STATE \textbf{Initialize:} Set initial antenna positions $\boldsymbol{t}_n^0$ and initialize the iteration counter $\ell=0$.  
\REPEAT
\STATE Compute the required parameters for the surrogate function in (31a), including $\psi_{m,n}$ derived in (29).
\STATE For each iteration, solve the problem (P3.n) for $N$ antennas using convex optimization methods obtain updated antenna positions $\boldsymbol{t}_n^{\ell+1}$.
\STATE Set $\ell=\ell+1$.
\UNTIL The fractional increase of (31a) between two consecutive iterations is below a threshold $\zeta$.
\end{algorithmic}
\label{alg1}
\end{algorithm}

\vspace{-6pt}
\section{Achievable Rate Derivation and Antenna Position Optimization under ZF Beamforming}
\vspace{-3pt}
In this section, we extend our analysis and design to the case with ZF beamforming. We first derive a lower bound on the ergodic rate under equal power allocation, leveraging statistical CSI and the structure of small-timescale ZF beamforming. These expressions are then utilized for MAs' position optimization in an antenna-wise manner, enhancing large-timescale performance.

\vspace{-16pt}
\subsection{Ergodic Rate Analysis with ZF beamforming}
\vspace{-5pt}
With ZF beamforming, multiuser interference is completely eliminated. Following the approach in [33, Theorem 4], the ZF transmit beamforming vector for user $m$ is expressed as
\begin{align}
&\!\!\!\boldsymbol{w}_{\text{ZF},m}^{}=\sqrt{p_m} \nonumber
\\
&\!\!\!\times \!\!\frac{( \boldsymbol{I}_N-\boldsymbol{B}_m\left( \boldsymbol{t} \right) ( \boldsymbol{B}_{m}^{H}\left( \boldsymbol{t} \right) \boldsymbol{B}_m\left( \boldsymbol{t} \right)) ^{-1}\boldsymbol{B}_{m}^{H}\left( \boldsymbol{t} \right) ) \boldsymbol{h}_m\left( \boldsymbol{t} \right)}{\lVert ( \boldsymbol{I}_N-\boldsymbol{B}_m\left( \boldsymbol{t} \right) ( \boldsymbol{B}_{m}^{H}\left( \boldsymbol{t} \right) \boldsymbol{B}_m\left( \boldsymbol{t} \right) ) ^{-1}\boldsymbol{B}_{m}^{H}\left( \boldsymbol{t} \right) ) \boldsymbol{h}_m\left( \boldsymbol{t} \right) \rVert},  \!
\end{align}
where $p_m$ represents the power equally allocated to user $m$, and $\boldsymbol{B}_m\left( \boldsymbol{t} \right)$ is defined as the projection matrix for user 
$m$'s channel, ensuring orthogonality to other users' channels
\begin{align}
\!\!\!\!\boldsymbol{B}_m\left( \boldsymbol{t} \right) \!=\!\left[ \boldsymbol{h}_1\left( \boldsymbol{t} \right) ,\dots ,\boldsymbol{h}_{m-1}\left( \boldsymbol{t} \right) ,\boldsymbol{h}_{m+1}\left( \boldsymbol{t} \right) ,\dots ,\boldsymbol{h}_M\left( \boldsymbol{t} \right) \right]. \!\!\!
\end{align}
Substituting (33) into (4) and applying equal power allocation
\begin{align}
p_m=p=\frac{P_\text{tot}}{M},
\end{align}
the resulting SNR at the receiver for user $m$ is given by 
\begin{align}
&\gamma_{\text{ZF},m}=\frac{p_m}{\sigma _{m}^{2}} \nonumber
\\
&\!\times \! \frac{| \boldsymbol{h}_{m}^{H}\left( \boldsymbol{t} \right) ( \boldsymbol{I}_N \!-\! \boldsymbol{B}_n\left( \boldsymbol{t} \right) ( \boldsymbol{B}_{m}^{H}\left( \boldsymbol{t} \right) \! \boldsymbol{B}_n\left( \boldsymbol{t} \right) ) ^{-1}\boldsymbol{B}_{m}^{H}\left( \boldsymbol{t} \right) ) \boldsymbol{h}_m\left( \boldsymbol{t} \right)|^2}{\lVert ( \boldsymbol{I}_N-\boldsymbol{B}_n\left( \boldsymbol{t} \right) ( \boldsymbol{B}_{m}^{H}\left( \boldsymbol{t} \right) \boldsymbol{B}_n\left( \boldsymbol{t} \right) ) ^{-1}\boldsymbol{B}_{m}^{H}\left( \boldsymbol{t} \right) ) \boldsymbol{h}_m\left( \boldsymbol{t} \right) \rVert ^2} \nonumber
\\
&\quad \quad \,\,  =\frac{p_m}{\sigma _{m}^{2}}\frac{1}{\big[ \big( \boldsymbol{H}^H\left( \boldsymbol{t} \right) \boldsymbol{H}\left( \boldsymbol{t} \right) \big) ^{-1} \big] _{mm}}, 
\end{align}
with $\boldsymbol{H}\left( \boldsymbol{t} \right) =\left[ \boldsymbol{h}_1\left( \boldsymbol{t} \right) ,\boldsymbol{h}_2\left( \boldsymbol{t} \right) ,\dots ,\boldsymbol{h}_M\left( \boldsymbol{t} \right) \right]$ represents the channel matrix.

With ZF beamforming, the antenna position optimization problem for ergodic sum rate maximization can be recast as
\begin{align}
\left( \text{P}4 \right) :\,\,\underset{\left\{ \boldsymbol{t}_n \right\}}{\max}&\,\,\mathbb{E}\left[ \log _2\left( 1+\gamma _{\text{ZF},m} \right) \right] 
\\
\text{s}.\text{t}.&\,\, (6\text{c}), (6\text{d}). \nonumber 
\end{align}
The exact expression for the ergodic rate under ZF beamforming is challenging to handle. To address this, we employ Jensen's inequality to derive a tight lower-bound approximation [33], which is expressed as\footnote{
Due to the nature of ZF beamforming that completely cancels multiuser interference, the proposed design cannot be directly applicable when $M>N$. However, we can adopt user grouping to divide and select smaller subsets of users. In this way, MA still offers extra degrees of freedom by shaping channel conditions and enhancing ergodic rate.}
\vspace{-6pt}
\begin{align}
\mathbb{E}\left\{ \log _2\left( 1+\gamma _{\text{ZF},m} \right) \right\} 
&\geq \log _2\Big( 1+\frac{p_m}{\sigma _{m}^{2}}\frac{\beta _m\left( N-M \right)}{\left[ \boldsymbol{\varSigma }^{-1}\left( \boldsymbol{t} \right) \right] _{mm}} \Big)  \nonumber
\\
&\triangleq \mathcal{R}_{\text{ZF},m}^{lb}(\boldsymbol{t}), 
\end{align}
where the expectation term is calculated based on statistical CSI, and $\boldsymbol{\varSigma }\left( \boldsymbol{t} \right)$ is defined as
\vspace{-6pt}
\begin{align}
\boldsymbol{\varSigma }\left( \boldsymbol{t} \right) =\boldsymbol{\varLambda}_1+\frac{1}{N}\boldsymbol{\varLambda}_2\boldsymbol{\bar{H}}_{}^{H}\left( \boldsymbol{t} \right) \boldsymbol{\bar{H}}\left( \boldsymbol{t} \right) \boldsymbol{\varLambda}_2,   
\end{align}
with the following deterministic terms and parameters
\begin{subequations}
\begin{align}
&\boldsymbol{\bar{H}}\left( \boldsymbol{t} \right) \triangleq \left[ \boldsymbol{\bar{h}}_1\left( \boldsymbol{t} \right) ,\boldsymbol{\bar{h}}_2\left( \boldsymbol{t} \right) ,\dots ,\boldsymbol{\bar{h}}_M\left( \boldsymbol{t} \right) \right], 
\\
&\boldsymbol{\varLambda}_1\triangleq \left( \boldsymbol{\varOmega }+\boldsymbol{I}_M \right) ^{-1},
\\
&\boldsymbol{\varLambda}_2\triangleq \left[ \boldsymbol{\varOmega \varLambda }_1 \right] ^{\frac{1}{2}},
\\
&\boldsymbol{\varOmega }\triangleq\mathrm{diag}\left( \left[ \kappa _1,\kappa _2,\dots ,\kappa _M \right] \right).
\end{align}
\end{subequations}

\textit{Remark 2:} From (37), (38), and (39), we can draw three key insights. First, under ZF beamforming in the two-timescale design, adjusting antenna positions does not enhance the ergodic rate when $M=1$. This is because $\boldsymbol{\varSigma }\left( \boldsymbol{t} \right)$ simplifies to a fixed scalar and $\mathcal{R}_{\text{ZF},m}^{lb}$ becomes a constant in this scenario. Second, if the deterministic LoS component for user $m$ is absent, i.e., $\kappa_m=0$, MAs cannot effectively reshape the channel, since $\left[ \boldsymbol{\varSigma }^{-1}\left( \boldsymbol{t} \right) \right] _{mm}$ in (37) becomes a fixed value of 1. Third, as $N$ becomes very large, the performance gain from MAs diminishes due to the channel hardening effect. In this case, an FPA system without antenna position tuning already achieves asymptotically orthogonal channels for different users, i.e., $\big|\frac{1}{N}\boldsymbol{\bar{H}}_j\left( \boldsymbol{t}_{\text{FPA}} \right) \boldsymbol{\bar{H}}_{m}^{H}\left( \boldsymbol{t}_{\text{FPA}} \right) \big|^2\rightarrow\boldsymbol{I}_N, \forall j,m$. Consequently, the properties of the ergodic rate in these scenarios are consistent with those described in Remark 1 for MRT beamforming.

\textit{Remark 3:} For the ZF beamforming that eliminates multiuser interference, the inverse channel term $[ ( \boldsymbol{H}^H\boldsymbol{H} ) ^{-1} ] _{mm}$ determines the SNR and thus the ergodic rate. From a statistical perspective, the sum-rate performance depends strongly on the ratio $\frac{N}{M}$ and the SNR [41]. As $M$ increases and approaches $N$, spatial degrees of freedom diminish, making the channel matrix ill-conditioned [42]. This amplifies $[ ( \boldsymbol{H}^H\boldsymbol{H})^{-1}]_{mm}$ and reduces the effective SNR for each user. Hence. FPA-ZF systems experience a marked decline in sum rate due to limited channel diversity and higher susceptibility to correlation. In the MA-enabled ZF system, antenna repositioning offers an additional degree of freedom to reshape $\boldsymbol{H}(\boldsymbol{t})$. Specifically, in the two-timescale framework, optimizing $\boldsymbol{t}$ to minimize $\left[ \boldsymbol{\varSigma }^{-1}\left( \boldsymbol{t} \right) \right] _{mm}$ can reduce off-diagonal elements in $\boldsymbol{H}^H\left( \boldsymbol{t} \right) \boldsymbol{H}\left( \boldsymbol{t} \right)$ (i.e., channel correlations), thus improving channel orthogonality. At the same time, the diagonal elements (overall channel power) remain largely preserved and thus increase the ergodic rate.

\vspace{-12pt}
\subsection{Antenna Position Optimization Based on AO Method}
\vspace{-3pt}
Despite having closed-form expressions for the ergodic rate in (37), the antenna positions remain intricately coupled in the implicit function $[ \boldsymbol{\varSigma }^{-1}\left( \boldsymbol{t} \right) ] _{mm}$, which depends primarily on the LoS components.
To tackle this complexity, we propose an element-wise antenna position optimization framework that leverages Woodbury's identity, the MM, and the SCA method. 

In this framework, we first derive a tractable expression for the achievable rate concerning the $n$-th antenna while keeping the positions of the other antennas fixed. Specifically, for the $n$-th antenna, we can express the effective channel matrix as
\begin{align}
\boldsymbol{\bar{H}}_{}^{H}\left( \boldsymbol{t}_n \right) \boldsymbol{\bar{H}}\left( \boldsymbol{t}_n \right) =\boldsymbol{\bar{g} }_n\left( \boldsymbol{t}_n \right) \boldsymbol{\bar{g}}_n^H\left( \boldsymbol{t}_n \right) +\boldsymbol{\varTheta }_{1,n},
\end{align}
where $\boldsymbol{\bar{g}}_n\left( \boldsymbol{t}_{n}^{} \right)$ is defined as the $n$-th row of $\boldsymbol{\bar{H}}$, i.e,
\begin{align}
\!\!\boldsymbol{\bar{g} }_n\left( \boldsymbol{t}_{n}^{} \right) \!\triangleq \!\big[ e^{j\frac{2\pi}{\lambda}\boldsymbol{t}_{n}^{T}\boldsymbol{a}_1},e^{j\frac{2\pi}{\lambda}\boldsymbol{t}_{n}^{T}\boldsymbol{a}_2},\dots ,e^{j\frac{2\pi}{\lambda}\boldsymbol{t}_{n}^{T}\boldsymbol{a}_M} \big] ^H \!\! \in \!\mathbb{C}^{M\times 1}, \!\!\!
\end{align}
and the matrix $\boldsymbol{\varTheta }_{1,n}$ is provided in (42) at the top of this page.
\begin{figure*}
\begin{align}
&\qquad\qquad \boldsymbol{\varTheta }_{1,n}=\left[ \begin{matrix}
	N-1&		\sum_{j=1,j\ne n}^N{e^{j\frac{2\pi}{\lambda}\boldsymbol{t}_{j}^{T}\left( \boldsymbol{a}_2-\boldsymbol{a}_1 \right)}}&		\cdots&		\sum_{j=1,j\ne n}^N{e^{j\frac{2\pi}{\lambda}\boldsymbol{t}_{j}^{T}\left( \boldsymbol{a}_M-\boldsymbol{a}_1 \right)}}\\
	\sum_{j=1,j\ne n}^N{e^{j\frac{2\pi}{\lambda}\boldsymbol{t}_{j}^{T}\left( \boldsymbol{a}_1-\boldsymbol{a}_2 \right)}}&		N-1&		\cdots&		\sum_{j=1,j\ne n}^N{e^{j\frac{2\pi}{\lambda}\boldsymbol{t}_{j}^{T}\left( \boldsymbol{a}_M-\boldsymbol{a}_2 \right)}}\\
	\vdots&		\vdots&		\ddots&		\vdots\\
	\sum_{j=1,j\ne n}^N{e^{j\frac{2\pi}{\lambda}\boldsymbol{t}_{j}^{T}\left( \boldsymbol{a}_1-\boldsymbol{a}_M \right)}}&		\sum_{j=1,j\ne n}^N{e^{j\frac{2\pi}{\lambda}\boldsymbol{t}_{j}^{T}\left( \boldsymbol{a}_2-\boldsymbol{a}_M \right)}}&		\cdots&		N-1\\
\end{matrix} \right]. 
\\
\vspace{-3pt}&\overline{\ \ \ \ \ \ \ \ \ \ \ \ \ \ \ \ \ \ \ \ \ \ \ \ \ \ \ \ \ \ \ \ \ \ \ \ \ \ \ \ \ \ \ \ \ \ \ \ \ \ \ \ \ \ \ \ \ \ \ \ \ \ \ \ \ \ \ \ \ \ \ \ \ \ \ \ \ \ \ \ \ \ \ \ \ \ \ \ \ \ \ \ \ \ \ \ \ \ \ \ \ \ \ \ \ \ \ \ \ \ \ \ \ \ \ \ \ \ \ \ \ \ \ \ \ \ \ \ \ \ \ \ \ \ \ \ \ \ \ \ \ \ \ \ } \nonumber
\end{align}
\vspace{-39pt}
\end{figure*}
Using these definitions, the matrix $\boldsymbol{\varSigma }\left( \boldsymbol{t}_n \right)$ is expressed as
\begin{align}
\boldsymbol{\varSigma }\left( \boldsymbol{t}_n \right) =\boldsymbol{\varTheta }_{2,n}+\frac{1}{N}\boldsymbol{\varLambda}_2\boldsymbol{\bar{g} }_n\left( \boldsymbol{t}_n \right) \boldsymbol{\bar{g} }_n^H\left( \boldsymbol{t}_n \right) \boldsymbol{\varLambda}_2, 
\end{align}
with $\boldsymbol{\varTheta }_{2,n}\triangleq\boldsymbol{\varLambda}_1+\frac{1}{N}\boldsymbol{\varLambda}_2\boldsymbol{\varTheta }_{1,n}\boldsymbol{\varLambda}_2$.
Given that $\boldsymbol{\varLambda}_1\succ 0$, $\boldsymbol{\varTheta }_{2,n}\succ 0$, $\boldsymbol{\varTheta }_{2,n}^{-1}\succ 0$, and $\boldsymbol{\varTheta}_{2,n}^{H}$ is Hermitian, 
we can apply Woodbury’s identity to rewrite $\boldsymbol{\varSigma }_{}^{-1}\left( \boldsymbol{t}_n \right)$ as follows [38]
\begin{subequations}
\begin{align}
\boldsymbol{\varSigma }_{}^{-1}\left( \boldsymbol{t}_n \right) &=\big( \boldsymbol{\varTheta }_{2,n}+\boldsymbol{\varLambda}_2\boldsymbol{\bar{g} }_n\left( \boldsymbol{t}_n \right) \frac{1}{N}\boldsymbol{\bar{g} }_n^H\left( \boldsymbol{t}_n \right) \boldsymbol{\varLambda}_2 \big) ^{-1}
\\
&=\boldsymbol{\varTheta }_{2,n}^{-1}-\frac{\boldsymbol{\varTheta }_{2,n}^{-1}\boldsymbol{\varLambda}_2\boldsymbol{\bar{g} }_n\left( \boldsymbol{t}_n \right) \boldsymbol{\bar{g} }_n^H\left( \boldsymbol{t}_n \right) \boldsymbol{\varLambda}_2\boldsymbol{\varTheta }_{2,n}^{-1}}{N+\boldsymbol{\bar{g} }_n^H\left( \boldsymbol{t}_n \right) \boldsymbol{\varLambda}_2\boldsymbol{\varTheta }_{2,n}^{-1}\boldsymbol{\varLambda}_2\boldsymbol{\bar{g} }_n\left( \boldsymbol{t}_n \right)}. \!\!
\end{align}
\end{subequations}

\textit{Remark 4:} From (44b), we derive the following inequality
\begin{align}
[\boldsymbol{\varSigma }_{}^{-1}\left( \boldsymbol{t}_n \right)]_{mm}\leq [\boldsymbol{\varTheta }_{2,n}^{-1}]_{mm}=[\boldsymbol{\varSigma }_{}^{-1}\left( \boldsymbol{t}_{\text{FPA},n} \right)]_{mm}.
\end{align}
The inequality holds strictly with MAs, whereas equality is achieved in the FPA case for the $n$-th antenna. This observation demonstrates that MAs offer a performance gain over FPAs by reducing the inverse channel matrix component $[\boldsymbol{\varSigma }_{}^{-1}\left( \boldsymbol{t}_n \right)]_{mm}$. Specifically, this reduction indicates a decrease in channel correlation between user 
$m$ and other users while maintaining the channel power gain for user $m$.

Then, $[ \boldsymbol{\varSigma }_{}^{-1}\left( \boldsymbol{t}_n \right) ] _{mm}$ can be expressed more compactly as
\begin{align}
&\left[ \boldsymbol{\varSigma }_{}^{-1}\left( \boldsymbol{t}_n \right) \right]_{mm} \nonumber
\\
&=\big[ \boldsymbol{\varTheta }_{2,n}^{-1}\big] _{mm}-\frac{\big[ \boldsymbol{\varTheta }_{2,n}^{-1}\boldsymbol{\varLambda}_2\boldsymbol{\bar{g} }_n\left( \boldsymbol{t}_n \right) \boldsymbol{\bar{g} }_n^H\left( \boldsymbol{t}_n \right) \boldsymbol{\varLambda}_2\boldsymbol{\varTheta }_{2,n}^{-1} \big] _{mm}}{N+\boldsymbol{\bar{g} }_n^H\left( \boldsymbol{t}_n \right) \boldsymbol{\varLambda}_2\boldsymbol{\varTheta }_{2,n}^{-1}\boldsymbol{\varLambda}_2\boldsymbol{\bar{g} }_n\left( \boldsymbol{t}_n \right)} \nonumber
\\
&=\frac{\boldsymbol{\bar{g} }_n^H\left( \boldsymbol{t}_n \right) \left[ \boldsymbol{\varTheta }_{2,n}^{-1} \right] _{mm}\boldsymbol{Y}_n\boldsymbol{\bar{g} }_n\left( \boldsymbol{t}_n \right)}{\boldsymbol{\bar{g} }_n^H\left( \boldsymbol{t}_n \right) \boldsymbol{Y}_n\boldsymbol{\bar{g} }_n\left( \boldsymbol{t}_n \right)}-\frac{\boldsymbol{l}_{n,m}^{H}\boldsymbol{\bar{g} }_n\left( \boldsymbol{t}_n \right) \boldsymbol{\bar{g} }_n^H\left( \boldsymbol{t}_n \right) \boldsymbol{l}_{n,m}^{}}{\boldsymbol{\bar{g} }_n^H\left( \boldsymbol{t}_n \right) \boldsymbol{Y}_n\boldsymbol{\bar{g} }_n\left( \boldsymbol{t}_n \right)} \nonumber
\\
&=\frac{\boldsymbol{\bar{g} }_n^H\left( \boldsymbol{t}_n \right) \big\{ \left[ \boldsymbol{\varTheta }_{2,n}^{-1} \right] _{mm}\boldsymbol{Y}_n-\boldsymbol{l}_{n,m}^{}\boldsymbol{l}_{n,m}^{H} \big\} \boldsymbol{\bar{g} }_n\left( \boldsymbol{t}_n \right)}{\boldsymbol{\bar{g} }_n^H\left( \boldsymbol{t}_n \right) \boldsymbol{Y}_n\boldsymbol{\bar{g} }_n\left( \boldsymbol{t}_n \right)} \nonumber
\\
&=\frac{\boldsymbol{\bar{g} }_n^H\left( \boldsymbol{t}_n \right) \boldsymbol{X}_{n,m}\boldsymbol{\bar{g} }_n\left( \boldsymbol{t}_n \right)}{\boldsymbol{\bar{g} }_n^H\left( \boldsymbol{t}_n \right) \boldsymbol{Y}_n\boldsymbol{\bar{g} }_n\left( \boldsymbol{t}_n \right)}.
\end{align}
where we define the following vectors and matrices
\begin{subequations}
\begin{align}
&\boldsymbol{Y}_n=\frac{N}{M}\boldsymbol{I}_M+\boldsymbol{\varLambda}_2\boldsymbol{\varTheta }_{2,n}^{-1}\boldsymbol{\varLambda}_2, 
\\
&\boldsymbol{l}_{n,m}^{H}=\left[ \boldsymbol{\varTheta }_{2,n}^{-1}\boldsymbol{\varLambda}_2 \right] _{\left( m,: \right)},
\\
&\boldsymbol{X}_{n,m}=\left[ \boldsymbol{\varTheta }_{2,n}^{-1} \right] _{mm}\boldsymbol{Y}_n-\boldsymbol{l}_{n,m}^{}\boldsymbol{l}_{n,m}^{H},
\end{align}
\end{subequations}
with $[ \cdot ] _{\left( m,: \right)}$ referred to the $m$-th row vector of a matrix.

Based on this, the lower bound on the ergodic rate for user $m$ is given by
\begin{align}
R_{\text{ZF},m}^{\text{lb,1}}\left( \boldsymbol{t}_n \right) =\log _2\Big( 1+\eta _m\frac{\boldsymbol{\bar{g} }_n^H\left( \boldsymbol{t}_n \right) \boldsymbol{Y}_n\boldsymbol{\bar{g} }_n\left( \boldsymbol{t}_n \right)}{\boldsymbol{\bar{g} }_n^H\left( \boldsymbol{t}_n \right) \boldsymbol{X}_{n,m}\boldsymbol{\bar{g} }_n\left( \boldsymbol{t}_n \right)} \Big), 
\end{align}
where we denote $\eta _m=\frac{p_m}{\sigma _{m}^{2}}\beta _m\left( N-M \right)$ for brevity.

The antenna position optimization problem for $\boldsymbol{t}_n$ with ZF beamforming can be reformulated as
\begin{align}
\left( \text{P}4.n \right): \,\, \underset{\boldsymbol{t}_n}{\max}& \,\,\sum_{m=1}^M{R_{\text{ZF},m}^{\text{lb,1}}\left( \boldsymbol{t}_n \right)}
\\
\text{s}.\text{t}.& \,\, ( 6\text{c}), ( 6\text{d}). \nonumber
\end{align}
Although the objective function is simplified, the fractional term $\frac{\boldsymbol{\bar{g} }_n^H\left( \boldsymbol{t}_n \right) \boldsymbol{Y}_n\boldsymbol{\bar{g} }_n\left( \boldsymbol{t}_n \right)}{\boldsymbol{\bar{g} }_n^H\left( \boldsymbol{t}_n \right) \boldsymbol{X}_{n,m}\boldsymbol{\bar{g} }_n\left( \boldsymbol{t}_n \right)}
$ remains neither concave nor convex with respect to $\boldsymbol{t}_{n}$, presenting challenges in constructing a global concave lower bound for the objective function.

\vspace{-12pt}
\subsection{Constructing a Surrogate Function via MM and SCA}
\vspace{-3pt}
To address the non-convexity of the objective function in (49), we construct a concave lower-bound surrogate function for $\boldsymbol{t}_{n}$ through a two-step process. This involves applying the MM method followed by the second-order Taylor expansion. The first step is summarized in the following lemma [39].

\textit{Lemma 1}: $R_{\text{ZF},m}^{\text{lb}}\left( \boldsymbol{t}_n \right)$ is lower bounded by
\begin{align}
&R_{\text{ZF},m}^{\text{lb},1}\left( \boldsymbol{t}_n \right) \ge R_{\text{ZF},m}^{\text{lb},2}\left( \boldsymbol{t}_{n}^{} \right)   \nonumber
\\
&\quad\quad\quad\quad\,\,=\log _2\left( 1+\eta _m f_{n,m}\left( \boldsymbol{\bar{g} }_n\left( \boldsymbol{t}_{n}^{} \right) \right) \right),
\end{align}
where the fractional term $\frac{\boldsymbol{\bar{g} }_n^H\left( \boldsymbol{t}_n \right) \boldsymbol{Y}_n\boldsymbol{\bar{g} }_n\left( \boldsymbol{t}_n \right)}{\boldsymbol{\bar{g} }_n^H\left( \boldsymbol{t}_n \right) \boldsymbol{X}_{n,m}\boldsymbol{\bar{g} }_n\left( \boldsymbol{t}_n \right)}
$ in (48) is replaced by the function $f_{n,m}\left( \boldsymbol{\bar{g} }_n\left( \boldsymbol{t}_{n}^{} \right) \right)$, defined as
\begin{align}
&\!\! \!f_{n,m}\left( \boldsymbol{\bar{g} }_n\left( \boldsymbol{t}_{n}^{} \right) \right) 
\nonumber
\\
&\!\!\!\triangleq \frac{2\text{Re}\big\{ \boldsymbol{\bar{g} }_n^H\left( \boldsymbol{t}_{n}^{\ell} \right) \boldsymbol{Y}_n\boldsymbol{\bar{g} }_n\left( \boldsymbol{t}_{n}^{} \right) \big\}}{\boldsymbol{\bar{g} }_n^H\left( \boldsymbol{t}_{n}^{\ell} \right) \boldsymbol{X}_{n,m}\boldsymbol{\bar{g} }_n\left( \boldsymbol{t}_{n}^{\ell} \right)}-\frac{\boldsymbol{\bar{g} }_n^H\left( \boldsymbol{t}_{n}^{\ell} \right) \boldsymbol{Y}_n\boldsymbol{\bar{g} }_n\left( \boldsymbol{t}_{n}^{\ell} \right)}{\big( \boldsymbol{\bar{g} }_n^H\left( \boldsymbol{t}_{n}^{\ell} \right) \boldsymbol{X}_{n,m}\boldsymbol{\bar{g} }_n\left( \boldsymbol{t}_{n}^{\ell} \right) \big) ^2} \nonumber
\\
&\,\,\times \!\big\{  2\lambda _{\max}\left( \boldsymbol{X}_{n,m} \right) M-\boldsymbol{\bar{g} }_n^H\left( \boldsymbol{t}_{n}^{\ell} \right) \boldsymbol{X}_{n,m}\boldsymbol{\bar{g} }_n\left( \boldsymbol{t}_{n}^{\ell} \right) \nonumber
\\
& \quad    +\!2\text{Re}\{ \boldsymbol{\bar{g} }_n^H\left( \boldsymbol{t}_{n}^{\ell} \right) \left[ \boldsymbol{X}_{n,m}\!-\!\!\lambda _{\max}\left( \boldsymbol{X}_{n,m} \right)\! \boldsymbol{I}\!_M \right] \boldsymbol{\bar{g} }_n\!\left( \boldsymbol{t}_n \right) \} \big\}. \!\!\!
\end{align}
\begin{proof}
The function $\log \left( 1+v \right) $ is a concavity-preserving transformation, so our task is to find a concave lower bound for $\frac{\boldsymbol{\bar{g} }_n^H\left( \boldsymbol{t}_n \right) \boldsymbol{Y}_n\boldsymbol{\bar{g} }_n\left( \boldsymbol{t}_n \right)}{\boldsymbol{\bar{g} }_n^H\left( \boldsymbol{t}_n \right) \boldsymbol{X}_{n,m}\boldsymbol{\bar{g} }_n\left( \boldsymbol{t}_n \right)}$. Define $r_{n,m}\left( \boldsymbol{t}_n \right) \triangleq\boldsymbol{\bar{g} }_n^H\left( \boldsymbol{t}_n \right) \boldsymbol{X}_{n,m}\boldsymbol{\bar{g} }_n\left( \boldsymbol{t}_n \right)$. Since $\boldsymbol{Y}_n$, defined in (46a), is positive definite, the function $\frac{\boldsymbol{\bar{g} }_n^H\left( \boldsymbol{t}_n \right) \boldsymbol{Y}_n\boldsymbol{\bar{g} }_n\left( \boldsymbol{t}_n \right)}{r_{n,m}^{}\left( \boldsymbol{t}_n \right)}$ is jointly convex with respect to $\boldsymbol{\bar{g} }_n\left( \boldsymbol{t}_n \right)$ and $r_{n,m}^{}\left( \boldsymbol{t}_n \right)$. Due to this convexity, the following inequality can be derived using the first-order Taylor expansion
\begin{align}
&\!\!\!\!\frac{\boldsymbol{\bar{g} }_n^H\left( \boldsymbol{t}_n \right) \boldsymbol{Y}_n\boldsymbol{\bar{g} }_n\left( \boldsymbol{t}_n \right)}{r_{n,m}^{}\left( \boldsymbol{t}_n \right)}\ge \frac{\boldsymbol{\bar{g} }_n^H\left( \boldsymbol{t}_{n}^{\ell} \right) \boldsymbol{Y}_n\boldsymbol{\bar{g} }_n\left( \boldsymbol{t}_{n}^{\ell} \right)}{r_{n,m}^{\ell}\left( \boldsymbol{t}_n \right)} \nonumber
\\
& +\frac{\partial \left( \frac{\boldsymbol{\bar{g} }_n^H\left( \boldsymbol{t}_n \right) \boldsymbol{Y}_n\boldsymbol{\bar{g} }_n\left( \boldsymbol{t}_n \right)}{r_{n,m}^{}\left( \boldsymbol{t}_n \right)} \right)}{\partial r_{n,m}^{}\left( \boldsymbol{t}_n \right)}\bigg|_{\boldsymbol{t}_n=\boldsymbol{t}_{n}^{\ell}}^{}\left( r_{n,m}^{}\left( \boldsymbol{t}_n \right) -r_{n,m}^{}\left( \boldsymbol{t}_{n}^{\ell} \right) \right) \nonumber
\\
& +\Big( \frac{\partial \big( \frac{\boldsymbol{\bar{g} }_n^H\left( \boldsymbol{t}_n \right) \boldsymbol{Y}_n\boldsymbol{\bar{g} }_n\left( \boldsymbol{t}_n \right)}{r_{n,m}^{}\left( \boldsymbol{t}_n \right)} \big)}{\partial \boldsymbol{\bar{g} }_n\left( \boldsymbol{t}_n \right)}\bigg|_{\boldsymbol{t}_n=\boldsymbol{t}_{n}^{\ell}}^{} \Big) ^H\left( \boldsymbol{\bar{g} }_n\left( \boldsymbol{t}_n \right) -\boldsymbol{\bar{g} }_n\left( \boldsymbol{t}_{n}^{\ell} \right) \right).\!\!\!\!
\end{align}
After calculating (52), the following inequalities can be established as a lower bound
\begin{align}
&\frac{\boldsymbol{\bar{g} }_n^H\left( \boldsymbol{t}_n \right) \boldsymbol{Y}_n\boldsymbol{\bar{g} }_n\left( \boldsymbol{t}_n \right)}{r_{n,m}^{}\left( \boldsymbol{t}_n \right)}\geq\frac{2\text{Re}\big\{ \boldsymbol{\bar{g} }_n^H\left( \boldsymbol{t}_{n}^{\ell} \right) \boldsymbol{Y}_n\boldsymbol{\bar{g} }_n\left( \boldsymbol{t}_{n}^{} \right) \big\}}{\boldsymbol{\bar{g} }_n^H\left( \boldsymbol{t}_{n}^{\ell} \right) \boldsymbol{X}_{n,m}\boldsymbol{\bar{g} }_n\left( \boldsymbol{t}_{n}^{\ell} \right)}\nonumber
\\
&\quad-\frac{\boldsymbol{\bar{g} }_n^H\left( \boldsymbol{t}_{n}^{\ell} \right) \boldsymbol{Y}_n\boldsymbol{\bar{g} }_n\left( \boldsymbol{t}_{n}^{\ell} \right)}{\big( \boldsymbol{\bar{g} }_n^H\left( \boldsymbol{t}_{n}^{\ell} \right) \boldsymbol{X}_{n,m}\boldsymbol{\bar{g} }_n\left( \boldsymbol{t}_{n}^{\ell} \right) \big) ^2}\boldsymbol{\bar{g} }_n^H\left( \boldsymbol{t}_n \right) \boldsymbol{X}_{n,m}\boldsymbol{\bar{g} }_n\left( \boldsymbol{t}_n \right).  \!\!
\end{align}
Note that the quadratic form in (53) is further bounded by
\begin{align}
&\!\!\boldsymbol{\bar{g} }_n^H\left( \boldsymbol{t}_n \right) \boldsymbol{X}_{n,m}\boldsymbol{\bar{g} }_n\left( \boldsymbol{t}_n \right) \nonumber
\\
&\!\!\le \boldsymbol{\bar{g} }_n^H\left( \boldsymbol{t}_n \right) \lambda _{\max}\left( \boldsymbol{X}_{n,m} \right) \boldsymbol{I}_M\boldsymbol{\bar{g} }\left( \boldsymbol{t}_n \right) \nonumber 
\\
&\!\!\quad +2\text{Re}\big\{ \boldsymbol{\bar{g} }_n^H\left( \boldsymbol{t}_{n}^{\ell} \right) \left[ \boldsymbol{X}_{n,m}-\lambda _{\max}\left( \boldsymbol{X}_{n,m} \right) \boldsymbol{I}_M \right] \boldsymbol{\bar{g} }_n\left( \boldsymbol{t}_n \right) \big\}  \nonumber
\\
&\!\!\quad +\boldsymbol{\bar{g} }_n^H\left( \boldsymbol{t}_{n}^{\ell} \right) \left[ \lambda _{\max}\left( \boldsymbol{X}_{n,m} \right) \boldsymbol{I}_M-\boldsymbol{X}_{n,m} \right] \boldsymbol{\bar{g} }_n\left( \boldsymbol{t}_{n}^{\ell} \right)  \nonumber
\\
&\!\!=2\lambda _{\max}\left( \boldsymbol{X}_{n,m} \right) M-\boldsymbol{\bar{g} }_n^H\left( \boldsymbol{t}_{n}^{\ell} \right) \boldsymbol{X}_{n,m}\boldsymbol{\bar{g} }_n\left( \boldsymbol{t}_{n}^{\ell} \right) \nonumber
\\
&\!\!\quad+2\text{Re}\big\{ \boldsymbol{\bar{g} }_n^H\left( \boldsymbol{t}_{n}^{\ell} \right) \! \left[ \boldsymbol{X}_{n,m}\!-\!\lambda _{\max}\left( \boldsymbol{X}_{n,m} \right) \boldsymbol{I}_M \right] \boldsymbol{\bar{g} }_n\left( \boldsymbol{t}_n \right) \!\big\}. \!\!\!
\end{align}
By substituting (53) and (54) into (48), we have (50).
\end{proof}

Using \textit{Lemma 1}, the lower-bound ergodic rate of user $m$ in (51) can be reformulated more concisely as
\begin{align}
R_{\text{ZF},m}^{\text{lb},2}\left( \boldsymbol{t}_{n}^{} \right) &=\log_2\big( 1+\eta _m\big( \chi _{n,m}^{\ell}+\text{Re}\left\{ \boldsymbol{q}_{n,m}^{T}\boldsymbol{\bar{g} }_n\left( \boldsymbol{t}_{n}^{} \right) \right\} \big) \big)  \nonumber
\\
& =\log_2\Big( 1+\eta _m\big( \chi _{n,m}^{\ell}+\sum_{u=1}^M{\left| q_{n,m,u}^{\ell} \right|} \nonumber
\\
&\quad\quad \quad \quad\quad  \times \cos ( \frac{2\pi}{\lambda}\boldsymbol{t}_{n}^{T}\boldsymbol{a}_u-\angle q_{n,m,u}^{\ell} ) \big) \Big),
\end{align}
where we define the constants specific to the optimization of $\boldsymbol{t}_{n}$ with the positions of the other antennas fixed as follows
\begin{align}
&\!\!\! \chi _{n,m}^{\ell}\triangleq -\frac{\boldsymbol{\bar{g} }_n^H\left( \boldsymbol{t}_{n}^{\ell} \right) \boldsymbol{Y}_n\boldsymbol{\bar{g} }_n\left( \boldsymbol{t}_{n}^{\ell} \right)}{\big( \boldsymbol{\bar{g} }_n^H\left( \boldsymbol{t}_{n}^{\ell} \right) \boldsymbol{X}_{n,m}\boldsymbol{\bar{g} }_n\left( \boldsymbol{t}_{n}^{\ell} \right) \big) ^2}
\Big( 2\lambda _{\max}\left( \boldsymbol{X}_{n,m} \right) M
\nonumber
\\
&\quad\quad\quad\quad\quad\quad\quad\quad\quad\quad-\boldsymbol{\bar{g} }_n^H\left( \boldsymbol{t}_{n}^{\ell} \right) \boldsymbol{X}_{n,m}\boldsymbol{\bar{g} }_n\left( \boldsymbol{t}_{n}^{\ell} \right) \Big), 
\\
&\!\!\! \boldsymbol{q}_{n,m}\triangleq \frac{2}{\boldsymbol{\bar{g} }_n^H\left( \boldsymbol{t}_{n}^{\ell} \right) \boldsymbol{X}_{n,m}\boldsymbol{\bar{g} }_n\left( \boldsymbol{t}_{n}^{\ell} \right)}\boldsymbol{\bar{g} }_n^H\left( \boldsymbol{t}_{n}^{\ell} \right) \Big(  \boldsymbol{Y}_n
\nonumber
\\
&-\frac{\boldsymbol{\bar{g} }_n^H\left( \boldsymbol{t}_{n}^{\ell} \right) \boldsymbol{Y}_n\boldsymbol{\bar{g} }_n\left( \boldsymbol{t}_{n}^{\ell} \right)}{\boldsymbol{\bar{g} }_n^H\left( \boldsymbol{t}_{n}^{\ell} \right) \boldsymbol{X}_{n,m}\boldsymbol{\bar{g} }_n\left( \boldsymbol{t}_{n}^{\ell} \right)}\!\left( \boldsymbol{X}_{n,m}\!-\!\lambda _{\max}\left( \boldsymbol{X}_{n,m} \right) \!\boldsymbol{I}\!_M \right) \!\Big). \!\!\!
\end{align}
Similar to problem (P2), we continue to address the non-convexity of the term $ \cos \big( \frac{2\pi}{\lambda}\boldsymbol{t}_{n}^{T}\boldsymbol{a}_m-\angle q_{n,m,u}^{\ell} \big)$ in (55) using the SCA technique. Given the structural similarity of the expressions, the SCA procedure mirrors the content outlined in Section III-C.
Specifically, by defining the non-convex non-concave function in (55) as
\vspace{-3pt}
\begin{align}
F_{n,m}\left( \boldsymbol{t}_{n}^{} \right) =\sum_{u=1}^M{\left| q_{n,m,u}^{\ell} \right|\cos \left(\frac{2\pi}{\lambda}\boldsymbol{t}_{n}^{T}\boldsymbol{a}_u-\angle q_{n,m,u}^{\ell} \right)},
\end{align}
the ergodic rate of user $m$ in (58) is further lower-bounded by
\vspace{-3pt}
\begin{align}
&R_{\text{ZF},m}^{\text{lb},2}\left( \boldsymbol{t}_{n}^{} \right) \ge R_{\text{ZF},m}^{\text{lb},3}\left( \boldsymbol{t}_{n}^{} \right) =\log _2\big( 1+\eta _m \big( \chi _{n,m}^{\ell}+F_{n,m}( \boldsymbol{t}_{n}^{\ell} )  \nonumber
\\
&+\nabla F_m( \boldsymbol{t}_{n}^{\ell} ) ^T \times( \boldsymbol{t}_{n}^{}-\boldsymbol{t}_{n}^{\ell} ) -\frac{\psi _{m,n}}{2}\| \boldsymbol{t}_{n}^{}-\boldsymbol{t}_{n}^{\ell} \| ^2 \big) \big),
\end{align}
where $\nabla ( F_{n,m})$ is the gradient of $F_{n,m}$, as provided in (60) at the top of the following page. 
\begin{figure*}
\begin{align}
&\nabla  F_{n,m}\left( \boldsymbol{t}_{n}^{\ell} \right) =\left[ \begin{array}{c}
	\frac{\partial F_{n,m}\left( \boldsymbol{t}_{n}^{} \right)}{\partial x_n}\Big|_{\boldsymbol{t}_{n}^{}=\boldsymbol{t}_{n}^{\ell}}^{}\\
	\frac{\partial F_{n,m}\left( \boldsymbol{t}_{n}^{} \right)}{\partial y_n}\Big|_{\boldsymbol{t}_{n}^{}=\boldsymbol{t}_{n}^{\ell}}^{}\\
\end{array} \right] 
=\left[ \begin{array}{c}
	-\frac{2\pi}{\lambda}\sum_{u=1}^M{\left| q_{n,m,u}^{\ell} \right|\cos \theta _{u}^{}\sin \phi _{u}^{}\sin \left( \frac{2\pi}{\lambda}\boldsymbol{t}_{n}^{\ell ,T}\boldsymbol{a}_u-\angle q_{n,m,u}^{\ell} \right)}\\
	-\frac{2\pi}{\lambda}\sum_{u=1}^M{\left| q_{n,m,u}^{\ell} \right|\sin \theta _{u}^{}\sin \left( \frac{2\pi}{\lambda}\boldsymbol{t}_{n}^{\ell ,T}\boldsymbol{a}_u-\angle q_{n,m,u}^{\ell} \right)}\\
\end{array} \right]. 
\\
&\overline{\ \ \ \ \ \ \ \ \ \ \ \ \ \ \ \ \ \ \ \ \ \ \ \ \ \ \ \ \ \ \ \ \ \ \ \ \ \ \ \ \ \ \ \ \ \ \ \ \ \ \ \ \ \ \ \ \ \ \ \ \ \ \ \ \ \ \ \ \ \ \ \ \ \ \ \ \ \ \ \ \ \ \ \ \ \ \ \ \ \ \ \ \ \ \ \ \ \ \ \ \ \ \ \ \ \ \ \ \ \ \ \ \ \ \ \ \ \ \ \ \ \ \ \ \ \ \ \ \ \ \ \ \ \ \ \ \ \ \ \ \ \ \ \ } \nonumber
\end{align}
\vspace{-36pt}
\end{figure*}
This concave lower bound is constructed by applying the second-order Taylor expansion 
\begin{align}
F_{n,m}\left( \boldsymbol{t}_{n}^{} \right) &\ge  F_{n,m}\left( \boldsymbol{t}_{n}^{\ell} \right) +\nabla  F_{n,m}\left( \boldsymbol{t}_{n}^{\ell} \right) ^T\left( \boldsymbol{t}_{n}^{}-\boldsymbol{t}_{n}^{\ell} \right) \nonumber
\\
&\quad-\frac{\xi _{m,n}}{2}\left( \boldsymbol{t}_{n}^{}-\boldsymbol{t}_{n}^{\ell} \right) ^T\left( \boldsymbol{t}_{n}^{}-\boldsymbol{t}_{n}^{\ell} \right) \nonumber
\\
&\triangleq  F_{n,m}^{\text{lb},\ell}\left( \boldsymbol{t}_{n}^{} \right),
\end{align}
where $\xi_{m,n}$ is a positive real number $\xi_{m,n}$ such that $\xi_{m,n}\boldsymbol{I}\succeq \nabla ^2F_m\left( \boldsymbol{t}_{n}^{} \right)$ and, similar to (29), can be given by 
\vspace{-3pt}
\begin{subequations}
\begin{align}
\xi _{m,n}&=\frac{4\pi ^2}{\lambda ^2} \epsilon _{\max}\left( \boldsymbol{\varXi} \right) 
\\
&=\frac{2\pi ^2}{\lambda ^2}\Big( \varXi _{11}+\varXi _{22}+\sqrt{\left( \varXi _{11}-\varXi _{22} \right) ^2+4\varXi _{12}^{2}} \Big), \!\!\!\!\!
\end{align}
\end{subequations}
with $\epsilon _{\max}(\cdot)$ denoting the maximum eigenvalue and the terms
\vspace{-15pt}
\begin{subequations}
\begin{align}
&\varXi _{11}=\sum_{u=1}^M{\left| q_{n,m,u}^{\ell} \right|\cos ^2\theta _{u}^{}\sin ^2\phi _{u}^{}},
\\
&\varXi _{12}^{}=\sum_{u=1}^M{\left| q_{n,m,u}^{\ell} \right|\left| \cos \theta _{u}^{}\sin \phi _{u}^{}\sin \theta _{u}^{} \right|},
\\
&\varXi _{22}=\sum_{u=1}^M{\left| q_{n,m,u}^{\ell} \right|\sin ^2\theta _{u}^{}}.
\end{align}
\end{subequations}

\vspace{-15pt}
\subsection{Overall Algorithm and Computational Complexity Analysis}
\vspace{-3pt}
The convex optimization problem for maximizing the lower bound of the ergodic sum rate can now be reformulated as
\begin{align}
\left( \text{P}5.n \right): \,\, \underset{\boldsymbol{t}_{n}^{}}{\max}&\,\,\sum_{m=1}^M{R_{\text{ZF},m}^{\text{lb},3}\left( \boldsymbol{t}_{n}^{} \right)}
\\
\text{s}.\text{t}. & \,\, (29\text{b}), (6\text{d}).  \nonumber
\end{align}
As a convex QCP, the optimal solution to this problem can be obtained using standard solvers (e.g., CVX [36]).

\begin{algorithm}[t]
\caption{AO-based optimization algorithm for (P4).}\label{alg:alg1}
\begin{algorithmic}[1]
\STATE \textbf{Initialize:} Set initial antenna positions $\boldsymbol{t}_n^0$ and initialize the iteration counter $\ell=0$.
\REPEAT
\STATE Compute the required parameters for the surrogate function in (64), including $\xi_{m,n}$ derived in (62).
\STATE For each iteration, solve the problem (P3.n) for $N$ antennas using convex optimization methods obtain updated antenna positions $\boldsymbol{t}_n^{\ell+1}$.
\STATE Set $\ell=\ell+1$.
\UNTIL The fractional increase of (65) between two consecutive iterations is below a threshold $\zeta$.
\end{algorithmic}
\label{alg1}
\end{algorithm}

The overall AO-based algorithm is detailed in Algorithm 2, which iteratively solves (P5.n) to provide a solution to the problem (P4). The algorithm guarantees convergence to a locally optimal solution, as the objective function value is non-decreasing in each iteration. 
Algorithm 2 optimizes the MA positions under ZF beamforming, extending the AO and SCA framework of MA-MRT with an additional MM step, which significantly increases complexity. The total computational complexity rises to $\mathcal{O}( L_2( N^{2.5}\log \frac{1}{\varepsilon}+NM^4))$, mainly originating from tcomputing the surrogate function that involves matrix inverses and eigenvalue calculations in $\mathcal{O}(M^3)$ for each user and each antenna in an AO iteration, totaling $\mathcal{O}(NM^4)$. Apart from this, the QCP solving step is identical to Algorithm 1. In the small timescale, beamforming vectors are computed using closed-form solutions once antenna positions are fixed. MRT beamforming requires $\mathcal{O}(MN)$ operations to compute the beamforming vectors based on channel coefficients, while the ZF beamforming involves a matrix inversion, costing $\mathcal{O}(M^3)$. These low-complexity operations are executed per channel realization, enabling efficient adaptation to real-time channel variations.

Although large-timescale optimization has a complexity of $\mathcal{O}( L_1( N^{2.5}\log \frac{1}{\varepsilon}+NM^2+MN^2 ))$ for MA-MRT and $\mathcal{O}( L_2( N^{2.5}\log \frac{1} {\varepsilon}+NM^4))$ for MA-ZF, for MA-ZF, its computational burden is alleviated by the temporal separation inherent in the two-timescale framework.
This framework improves practical feasibility by distributing computational tasks across different timescales, balancing performance optimization with efficiency. Specifically, antenna position optimization operates on a large timescale (e.g., every 10 seconds), covering multiple time slots where user distributions and channel statistics remain relatively stable. Meanwhile, small-timescale operations (e.g., every 10 milliseconds) manage real-time channel variations through conventional channel estimation and beamforming. By significantly reducing the frequency of computationally intensive antenna position updates, this separation makes the proposed MA-enabled system both efficient and scalable.

\vspace{-15pt}
\section{Ergodic Rate with Spatial Correlation}
\vspace{-9pt}
In this section, we justify our modeling assumptions for the NLoS channel components by extending the analysis beyond the simplified i.i.d. framework. We derive more accurate closed-form expressions for the ergodic rate in the MA-enabled MU-MIMO system under two-timescale design, incorporating a spatial correlation matrix $\boldsymbol{S}(\boldsymbol{t})$ that captures the covariance structure of the NLoS components. These derivations are provided for both MRT and ZF beamforming. 

With the spatial correlation matrix $\boldsymbol{S}(\boldsymbol{t})$, the Rician channel model is restated as follows
\begin{align}
\boldsymbol{h}_m\left( \boldsymbol{t} \right) =\sqrt{\frac{\kappa _m\beta _m}{\kappa _m+1}}\boldsymbol{\bar{h}}_m\left( \boldsymbol{t} \right) +\sqrt{\frac{\beta _m}{\kappa _m+1}}\boldsymbol{S}^{\frac{1}{2}}\left( \boldsymbol{t} \right) \boldsymbol{u}_m,
\end{align}
where $\boldsymbol{u}_m \sim \mathcal{C}\mathcal{N}\left( 0,\boldsymbol{I} \right)$, and $\boldsymbol{S}\left( \boldsymbol{t} \right)$ has the features $\boldsymbol{S}^{\frac{1}{2}}\left( \boldsymbol{t} \right) =\boldsymbol{S}^{\frac{1}{2}}\left( \boldsymbol{t} \right) ^H$ and$ \boldsymbol{S}^{\frac{1}{2}}\boldsymbol{S}^{\frac{1}{2}}\left( \boldsymbol{t} \right) =\boldsymbol{S}\left( \boldsymbol{t} \right)$.
Based on the approximation in [33, Lemma 1] and (12), after calculating the expectation terms, the ergodic rate for user $m$ with MRT beamforming is now expressed at (13) provided at the top of the next page.
\begin{figure*}
\begin{align}
& \!\!R_{\text{MRT},m}^{app}\!=\!\log _2\Bigg( \!1\!+\!\frac{\beta _{m}^{2}\Big( \frac{2\kappa _{m}^{}\boldsymbol{\bar{h}}_{m}^{H}\left( \boldsymbol{t} \right) \boldsymbol{S}\left( \boldsymbol{t} \right) \boldsymbol{\bar{h}}_m\left( \boldsymbol{t} \right) +\text{tr}\left( \boldsymbol{S}^2\left( \boldsymbol{t} \right) \right)}{\left( \kappa _m+1 \right) ^2}+N^2 \Big)}{\sum_{j=1,j\ne m}^M{\beta _{m}^{}\beta _{j}^{}\frac{\kappa _m\kappa _j\left| \boldsymbol{\bar{h}}_j\left( \boldsymbol{t} \right) \boldsymbol{\bar{h}}_{m}^{H}\left( \boldsymbol{t} \right) \right|^2+\kappa _m\boldsymbol{\bar{h}}_{m}^{H}\left( \boldsymbol{t} \right) \boldsymbol{S}\left( \boldsymbol{t} \right) \boldsymbol{\bar{h}}_{m}^{}\left( \boldsymbol{t} \right) +\kappa _j\boldsymbol{\bar{h}}_{j}^{H}\left( \boldsymbol{t} \right) \boldsymbol{S}\left( \boldsymbol{t} \right) \boldsymbol{\bar{h}}_{j}^{}\left( \boldsymbol{t} \right) \left. +\text{tr}\left( \boldsymbol{S}^2\left( \boldsymbol{t} \right) \right) \right)}{\left( \kappa _m+1 \right) \left( \kappa _j+1 \right)} \!+ \!\frac{\sigma _{m}^{2}}{P}N \!\sum_{m=1}^M{\beta _{m}^{}}}}\! \Bigg)\!. \!\!\!
\\
&\! \! \overline{\ \ \ \ \ \ \ \ \ \ \ \ \ \ \ \ \ \ \ \ \ \ \ \ \ \ \ \ \ \ \ \ \ \ \ \ \ \ \ \ \ \ \ \ \ \ \ \ \ \ \ \ \ \ \ \ \ \ \ \ \ \ \ \ \ \ \ \ \ \ \ \ \ \ \ \ \ \ \ \ \ \ \ \ \ \ \ \ \ \ \ \ \ \ \ \ \ \ \ \ \ \ \ \ \ \ \ \ \ \ \ \ \ \ \ \ \ \ \ \ \ \ \ \ \ \ \ \ \ \ \ \ \ \ \ \ \ \ \ \ \ \ \ \ \ } \nonumber
\end{align}
\vspace{-36pt}
\end{figure*}
For ZF beamforming, following (35) and (37), to further derive the closed-form ergodic rate expression with spatial correlation matrix, we first rewrite the channel matrix in a compact form
\begin{align}
\boldsymbol{H}(\boldsymbol{t}) =\boldsymbol{\bar{H}}( \boldsymbol{t} ) [\boldsymbol{\varOmega}( \boldsymbol{\varOmega }+\boldsymbol{I}_M ) ^{-1} ] ^{\frac{1}{2}}+\boldsymbol{S}^{\frac{1}{2}}( \boldsymbol{t} ) \boldsymbol{U} ( \boldsymbol{\varOmega }+\boldsymbol{I}_M) ^{-\frac{1}{2}},
\end{align}
where $\boldsymbol{U}=[\boldsymbol{u}_1,\dots,\boldsymbol{u}_M]$ is the Gaussian random variable matrix for NLoS components. Accordingly, $\boldsymbol{H}^H\left( \boldsymbol{t} \right) \boldsymbol{H}\left( \boldsymbol{t} \right)$ follows a non-central Wishart distribution [40],
\begin{align}
\boldsymbol{H}^H\left( \boldsymbol{t} \right) \boldsymbol{H}\left( \boldsymbol{t} \right) \sim \mathcal{W}_M\left( N,\boldsymbol{P},\boldsymbol{\varSigma } \right), 
\end{align}
with $N$ degrees of freedom, covariance matrix $\boldsymbol{\varSigma }$, and matrix of noncentrality parameters $\boldsymbol{P}$
\begin{align}
&\boldsymbol{\varSigma }\triangleq \left( \boldsymbol{\varOmega }+\boldsymbol{I}_M \right) ^{-\frac{1}{2}}\boldsymbol{S}\left( \boldsymbol{t} \right) \left( \boldsymbol{\varOmega }+\boldsymbol{I}_M \right) ^{-\frac{1}{2}},
\\
&\boldsymbol{P}\triangleq \boldsymbol{\varSigma }^{-1} [ \boldsymbol{\varOmega } ( \boldsymbol{\varOmega }+\boldsymbol{I}_M) ^{-1} ] ^{\frac{1}{2}}\boldsymbol{\bar{H}}^H ( \boldsymbol{t}) \boldsymbol{\bar{H}}( \boldsymbol{t}) [ \boldsymbol{\varOmega }( \boldsymbol{\varOmega }+\boldsymbol{I}_M ) ^{-1} ] ^{\frac{1}{2}}.
\end{align}
Now, we can approximate $\boldsymbol{H}^H\left( \boldsymbol{t} \right) \boldsymbol{H}\left( \boldsymbol{t} \right)$ by a central Wishart distribution with covariance matrix
\begin{align}
&\!\!\boldsymbol{\hat{\varSigma}}= ( \boldsymbol{\varOmega }+\boldsymbol{I}_M ) ^{-\frac{1}{2}}\boldsymbol{S} (\boldsymbol{t}) ( \boldsymbol{\varOmega }+\boldsymbol{I}_M ) ^{-\frac{1}{2}} \nonumber
\\
&\!\!+\!\frac{1}{N}[ \boldsymbol{\varOmega }( \boldsymbol{\varOmega }+\boldsymbol{I}_M ) ^{-1} ] ^{\frac{1}{2}}\boldsymbol{\bar{H}}^H ( \boldsymbol{t} ) \boldsymbol{\bar{H}}( \boldsymbol{t} )[ \boldsymbol{\varOmega }\left( \boldsymbol{\varOmega }+\boldsymbol{I}_M \right) ^{-1} ] ^{\frac{1}{2}}.\!\!\!
\end{align}
and the lower bound ergodic rate for user $m$ is expressed as
\begin{align}
\mathcal{R}_{\text{ZF},m}^{lb}(\boldsymbol{t}) = \log _2\big( 1+\frac{p_m}{\sigma _{m}^{2}}\frac{\beta _m\left( N-M \right)}{\left[ \boldsymbol{\hat{\varSigma} }^{-1}\left( \boldsymbol{t} \right) \right] _{mm}} \big). 
\end{align}

Using an identity covariance matrix can sometimes oversimplify the channel model, especially if antennas could be positioned arbitrarily close (e.g., $\lambda/10$). However, our design explicitly avoids such scenarios through the imposed \textit{minimum antenna spacing} constraint to justify this simplification. Specifically, we enforce that the inter-antenna spacing is always greater than $\lambda/2$ as shown in the constraint (6c). Under such configurations, the spatial correlation among different antennas is significantly reduced, and from the physics standpoint, the Rayleigh NLoS components across the array would not experience highly correlated scattering. In the simulation results, we also evaluated the performance deviation between the i.i.d. assumption and the correlated case. The ergodic rates containing the spatial correlation are compared with the simplified ones. Simulations with Bessel correlation functions show that the deviation remains limited and not significant with the minimum antenna spacing $\lambda/2$. In our simulation study, we evaluate the ergodic rate expressions that account for spatial correlation with those obtained under the simplified i.i.d. assumption, demonstrate that with a minimum antenna spacing, the deviation can be minimal and not significant.

\setlength{\textfloatsep}{12pt}
\vspace{-9pt}
\section{Simulation Results}
\vspace{-3pt}
This section provides numerical results to validate the effectiveness of the proposed two-timescale MA-enabled transmission design and our theoretical findings. In the simulations, users are randomly distributed around the BS, with their distances $d_m$ (in meters) uniformly sampled in the range of [50, 70]. 
The Rician fading channel model is employed with a common Rician factor $\kappa_m=\kappa,\forall m$ for all BS-user channels, providing consistent LoS conditions across users. The large-scale fading factor for each user $\beta_m$ is modeled as $\beta_m=\beta_0 d_m^{-\alpha}$, where $\beta_0 = -40$dB represents the reference average channel power gain at 1m, and $\alpha= 2.8$ denotes the path-loss exponent. 
Based on the user distribution, the elevation and azimuth angles of AoD and AoA for each channel path are uniformly distributed within $\left[ -\frac{\pi}{2},\frac{\pi}{2} \right]$. The movable regions for the transmit MAs are set as $\mathcal{C}=\left[ -\frac{N_rA\lambda}{2},\frac{N_rA\lambda}{2} \right] \times \left[ -\frac{N_cA\lambda}{2},\frac{N_cA\lambda}{2} \right]$, which expand adaptively with the number of antennas. Additionally, the minimum antenna spacing is constrained by $D_\text{min} = \frac{\lambda}{2}$, and the noise power is fixed at $\sigma^2_m = -80$dBm. The primary antenna-user configuration considered is $N=6,M=5$, with Rician factors $\kappa=6$ or $\kappa=100$, representing moderate and strong LoS conditions, respectively. The maximum transmit power $P_\text{tot}$ is set to 1W, and the region size parameter $A$ is set to 2 unless otherwise stated. Further simulation-specific parameters are outlined in the results section.

\setlength{\abovecaptionskip}{-3pt}

\begin{figure}[!t]
\centering
\includegraphics[width=2.4in,height=1.8in]{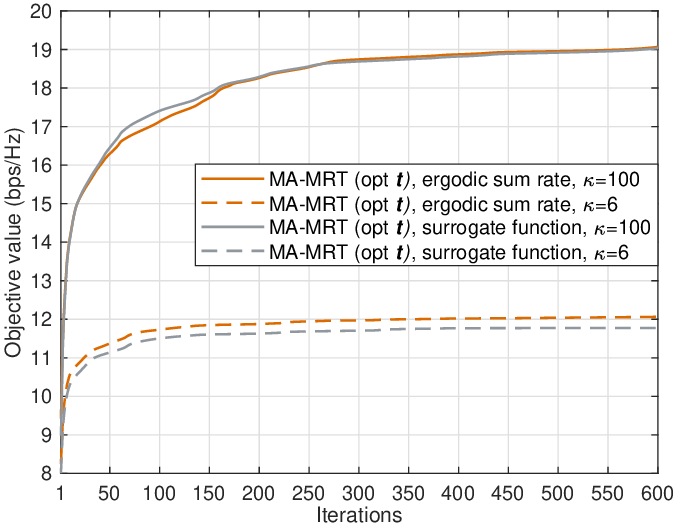}
\caption{The convergence of the proposed MA-MRT Algorithm 1.} 
\vspace{-12pt}
\end{figure}
\begin{figure}[!t]
\centering
\includegraphics[width=2.4in,height=1.8in]{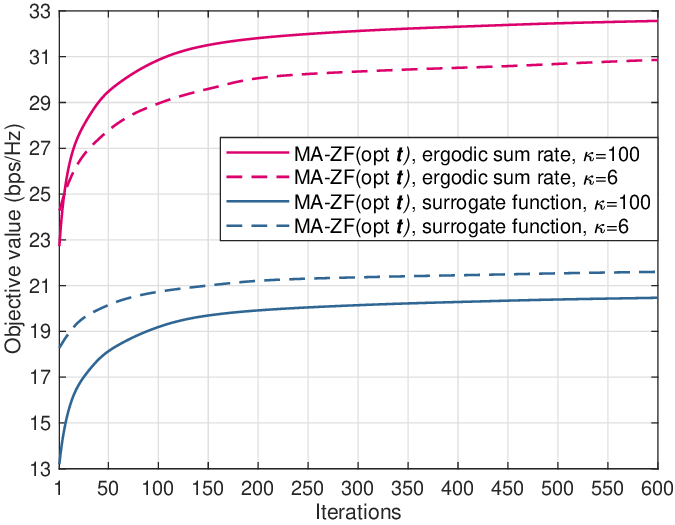}
\caption{The convergence of the proposed MA-ZF Algorithm 2.}
\vspace{-12pt}
\end{figure}

In the following, we evaluate the performance of the proposed schemes and compare them to the following benchmarks 
\begin{enumerate}
\item \textbf{MA with ZF beamforming}: Denoted as ``MA-ZF (opt $\boldsymbol{t}$)", this scheme uses ZF beamforming with fixed power allocation  (35), and the antenna positions are optimized using Algorithm 2, leveraging two-timescale CSI.
\item \textbf{MA with MRT beamforming}: Denoted by ``MA-MRT (opt $\boldsymbol{t}$)", this scheme employs simple MRT beamforming with power allocation (10), and the antenna positions are optimized via Algorithm 1, also using two-timescale CSI.
\item \textbf{FPA with ZF beamforming and power allocation}: ``FPA-ZF (opt $\boldsymbol{p}$)" uses ZF beamforming with optimized power allocation, leveraging instantaneous CSI.
\item \textbf{FPA with MRT beamforming and power allocation}: ``FPA-MRT (opt $\boldsymbol{p}$)", employs MRT beamforming with optimized power allocation based on instantaneous CSI.
\item \textbf{FPA with optimal beamforming}: ``FPA-OPT (opt $\boldsymbol{W}$)" applies optimal adaptive beamforming to the FPA system, using instantaneous CSI for performance comparison\footnote{In ``FPA-OPT (opt W)," the beamforming vectors for a fixed antenna configuration are iteratively optimized to maximize the MU-MIMO sum rate by reformulating the non-convex problem through fractional programming with a quadratic transform}.
\end{enumerate}

\vspace{-3pt}
Figs. 3 and 4 show the convergence of the proposed MA-MRT (Algorithm 1) and MA-ZF (Algorithm 2) schemes with $N=6,M=5$. Both algorithms optimize surrogate functions to approximate the ergodic sum rate, with a predetermined fraction increase threshold $\zeta=0.5\times10^{-4}$.
Both algorithms exhibit efficient and stable convergence, where the ergodic sum rate increases steadily as iterations progress, especially under strong LoS conditions.
MA-MRT converges within 450 iterations under both high ($\kappa=100$) and moderate ($\kappa=6$) Rician factors, with a minor gap between the surrogate and actual values. MA-ZF shows more steady and significant performance gains and typically converges within 500 iterations, with the surrogate function acting as a lower bound to the ergodic sum rate. 
These results demonstrate the effectiveness of the proposed methods in optimizing the ergodic sum rate under our proposed two-timescale framework.

\begin{figure}[t]
\centering
\includegraphics[width=2.4in,height=1.8in]{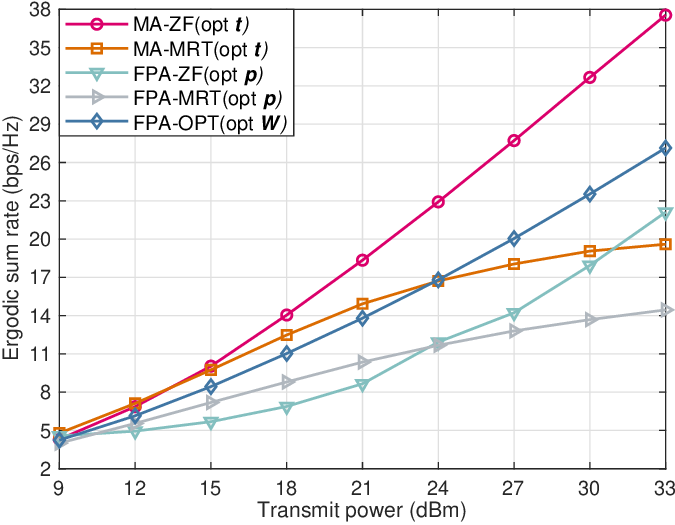}
\caption{Ergodic sum rate versus maximum transmit power with $N=6,M=5$ and $\kappa=100$.} 
\vspace{-12pt}
\end{figure}
\begin{figure}[t]
\centering
\includegraphics[width=2.4in,height=1.8in]{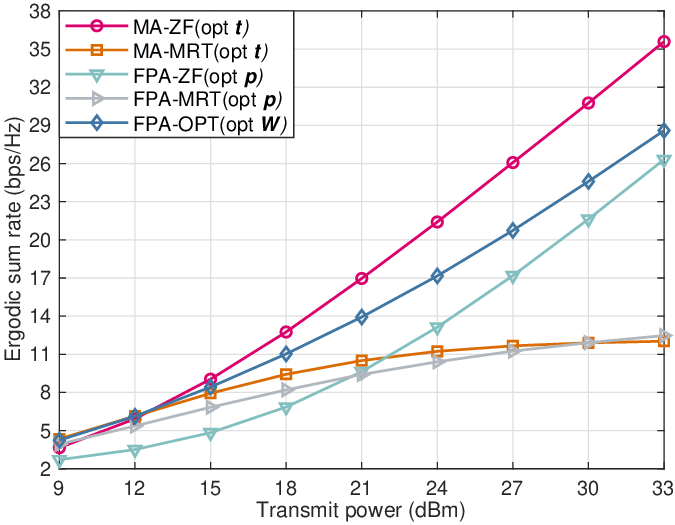}
\caption{Ergodic sum rate versus maximum transmit power with $N=6,M=5$ and $\kappa=6$.} 
\vspace{-9pt}
\end{figure}

\begin{figure}[t]
\centering
\includegraphics[width=2.4in,height=1.8in]{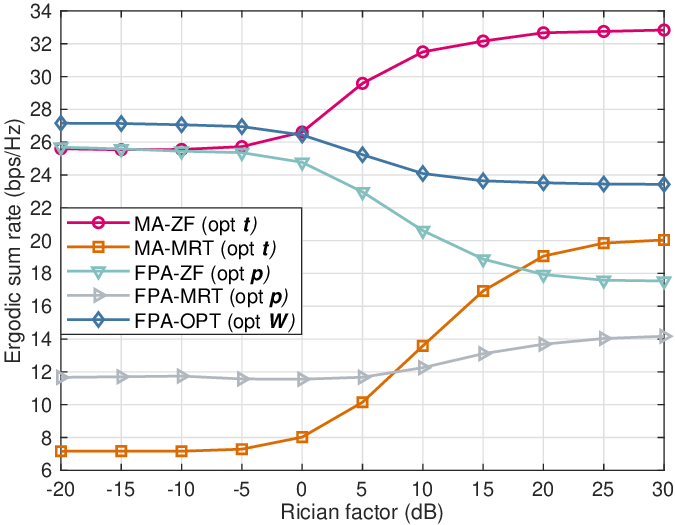}
\caption{Ergodic sum rate versus Rician factor with $N=6,M=5$.} 
\vspace{-9pt}
\end{figure}
\begin{figure}[t]
\centering
\includegraphics[width=2.4in,height=1.8in]{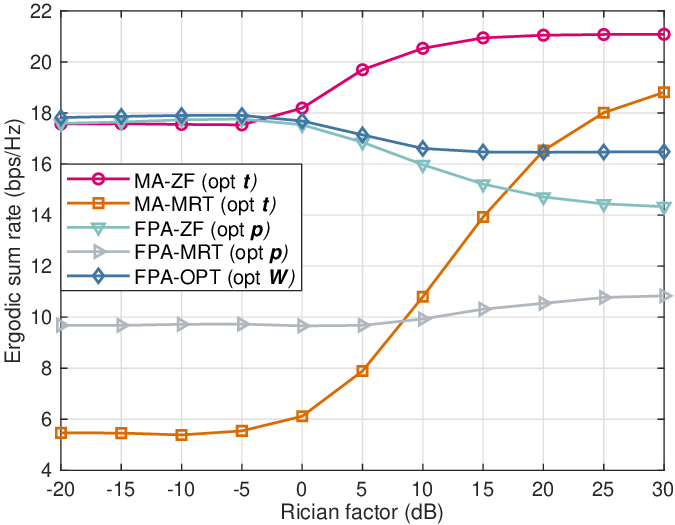}
\caption{Ergodic sum rate versus Rician factor with $N=4,M=3$.} 
\vspace{-9pt}
\end{figure}

Figs. 5 and 6 evaluate the ergodic sum rate performance of various schemes across different transmit power levels with $\kappa=100$ and $\kappa=6$, respectively. 
In both figures, MA-ZF consistently outperforms the other approaches across all transmit power levels except for the very low power scenario $P_\text{tot}=9dBm$, effectively exploiting spatial degrees of freedom provided by antenna repositioning. 
In contrast, MA-MRT exhibits a significantly lower ergodic sum rate compared to MA-ZF, highlighting the limitations of MRT beamforming in managing multiuser interference. In high transmit power regimes (above 15 dBm) under weaker LoS conditions ($\kappa$=6), MA-MRT lags behind FPA-MRT mainly because the stronger random NLoS component diminishes gains from antenna repositioning and MA-MRT’s simplistic power allocation is less adaptive than FPA-MRT’s fractional programming approach. A key observation is the impact of the Rician factor on system performance. As the Rician factor decreases from $\kappa=100$ to $\kappa=6$ transitioning from strong to weaker LoS conditions, there is a noticeable decline in performance for MA-enabled schemes, particularly MA-MRT, due to the reduced availability of deterministic channel components. Encouragingly, MA-ZF maintains robustness across both strong and moderate LoS scenarios.
In summary, MA-enabled systems exhibit superior or comparable performance to FPA-based systems, with MA-ZF consistently outperforming MA-MRT. These results highlight the performance gains achieved by optimizing antenna positions, especially in strong LoS environments. Nonetheless, it is important to note that MAs alone cannot replace well-designed transmit beamforming; instead, they work in synergy with beamforming to maximize overall system performance.

For a clearer investigation into the impact of Rician factors, Figs. 7 and 8 depict the ergodic sum rate as a function of the Rician factor $\kappa$. Across all scenarios, MA-enabled schemes benefit significantly from increasing $\kappa$, aligning with expectations under the two-timescale design. As the channel becomes more deterministic, the optimization potential of the proposed scheme expands.
Except in weak LoS conditions where $\kappa \leq 0$, MA-ZF consistently achieves the highest ergodic sum rate, even surpassing the performance of optimal beamforming. Remarkably, despite using equal power allocation and being a sub-optimal beamforming scheme, MA-ZF demonstrates the significant performance potential of MAs in moderate to strong LoS conditions, which underscores the importance of wireless channel reshaping and fully exploiting spatial degrees of freedom enabled by MAs.
MA-MRT also sees considerable improvement as $\kappa$ increases, benefiting from stronger deterministic LoS components. When $\kappa$ approaches 30 dB, MA-MRT reaches performance levels comparable to FPA-ZF and even FPA-OPT, especially with fewer users ($N=4,M=3$). However, MA-MRT remains inferior to MA-ZF across all values of $\kappa$, again highlighting the limitations of MRT beamforming in scenarios where interference suppression is crucial, such as systems with high user density ($N=6,M=5$).

\begin{figure}[t]
\centering
\includegraphics[width=2.4in,height=1.8in]{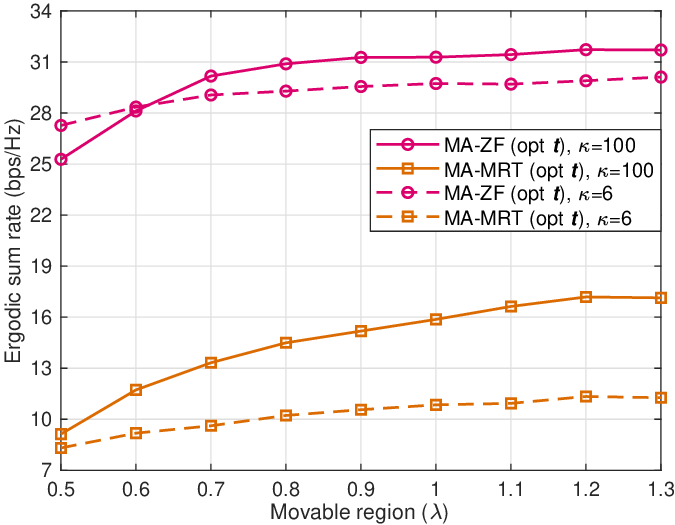}
\caption{Ergodic sum rate versus movable region size with $N=6,M=5$.} 
\vspace{-12pt}
\end{figure}
\begin{figure}[t]
\centering
\includegraphics[width=2.4in,height=1.8in]{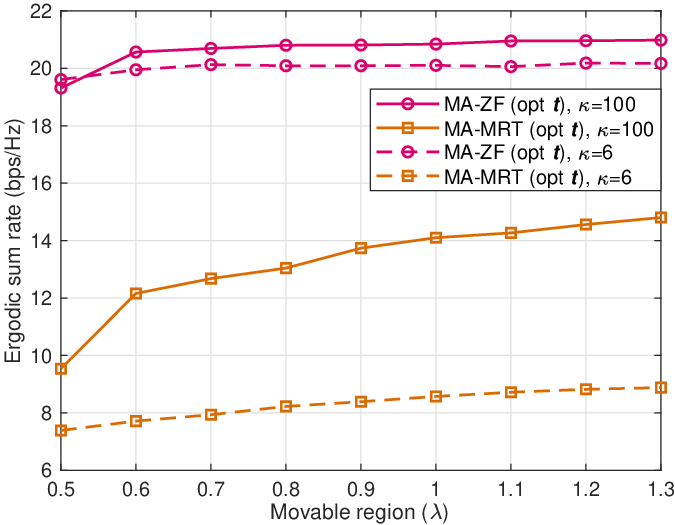}
\caption{Ergodic sum rate versus movable region size with $N=4,M=3$.} 
\vspace{-8pt}
\end{figure}

Figs. 9 and 10 illustrate the impact of the movable region size $A$ on the ergodic sum rate for the proposed schemes under different user-antenna configurations and Rician factors. As the region expands, both MA-ZF and MA-MRT benefit from increased flexibility in position tuning, leading to steady improvements in performance.
However, the performance gains begin to taper off beyond a certain region size, especially for MA-ZF, indicating that further enlarging the region offers diminishing returns. This suggests that the algorithm may have already converged to a locally optimal solution, and additional spatial flexibility provides limited benefit. In contrast, MA-MRT continues to see more pronounced improvements as the region expands, particularly under higher Rician factors. This indicates that the MA-based scheme with MRT beamforming requires a larger movable region to effectively mitigate interference and optimize performance. In other words, these results also imply that MA-ZF can exploit channel variation and achieve near-optimal performance more efficiently with a relatively smaller region, making it more effective for practical applications with constrained antenna movement.

\begin{figure}[t]
\centering
\includegraphics[width=2.4in,height=1.8in]{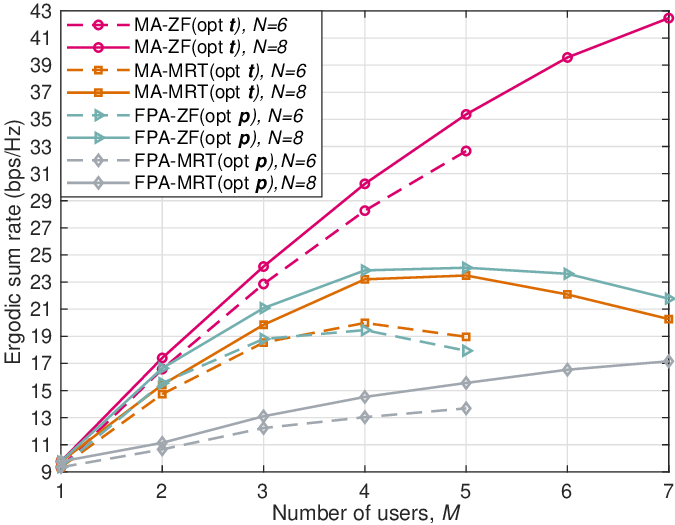}
\caption{Ergodic sum rate as a function of the number of users $M$ for MA and FPA configurations $N=6$ and $N=8$ with $\kappa=100$.}
\vspace{-12pt}
\end{figure}
\begin{figure}[!t]
\centering
\includegraphics[width=2.4in,height=1.8in]{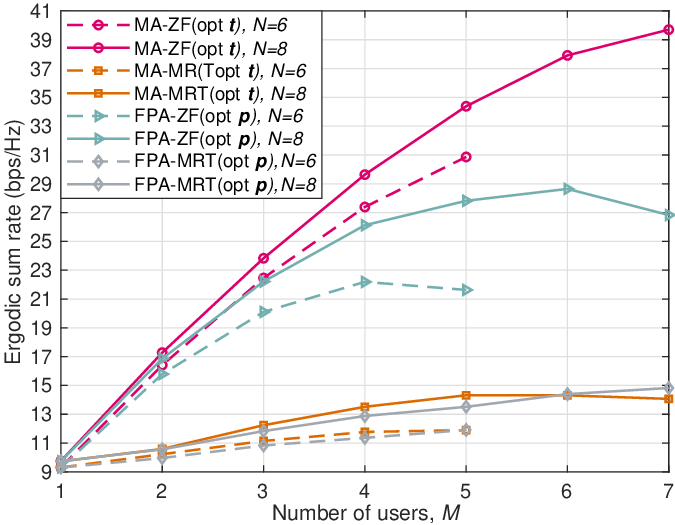}
\caption{Ergodic sum rate as a function of the number of users $M$ for MA and FPA configurations $N=6$ and $N=8$ with $\kappa=6$.}
\vspace{-9pt}
\end{figure}
Finally, Figs. 11 and 12 explore the relationship between the number of users $M$ and the ergodic sum rate for various MA and FPA configurations under ZF and MRT beamforming, with $N=6$ and $N=8$, respectively.
From both figures, it is observed that MA-ZF consistently delivers performance gains as the number of users increases, maintaining superiority over other schemes, particularly in scenarios with more users. In contrast, MA-MRT shows a gradual increase in performance but struggles with multiuser interference, leading to a decline in performance as the number of users approaches the number of antennas.
Under high Rician factor conditions, MA-MRT achieves sum rates comparable to FPA-ZF and notably outperforms FPA-MRT. Given its low complexity, MA-MRT offers a promising alternative to sophisticated beamforming designs in FPA systems.
In low Rician factor environments, the overall performance trends remain similar, but the rates drop due to the reduced deterministic LoS component, with MA-MRT being particularly impacted, rendering it almost ineffective. Despite this, MA-ZF continues to deliver superior performance as the number of users increases, showcasing its robustness in both strong and moderate LoS conditions.
Moreover, comparing the results for $N=6$ and $N=8$, the addition of antennas boosts sum rates across all configurations, with the improvement being more pronounced for MA-ZF. This highlights MA-ZF's ability to leverage spatial diversity for effective multiuser interference management.

Figs. 13-16 evaluate the impact of different NLoS modeling approaches, that is, spatially correlated vs. simplified i.i.d., on the ergodic sum rate, where the i.i.d. assumption is imposed with a minimum antenna spacing of $\lambda/2$.
\begin{figure}[!t]
\centering
\includegraphics[width=2.4in,height=1.8in]{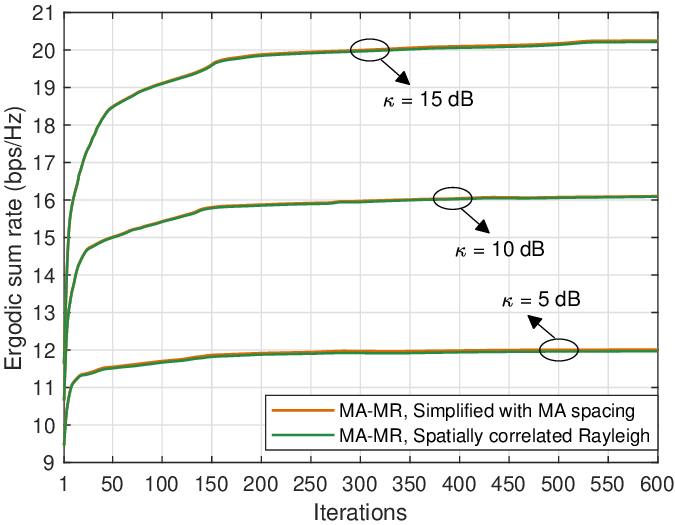}
\caption{Ergodic sum rate evaluation for MA-MRT with different NLoS models ($N=8,M=5$).} 
\vspace{-9pt}
\end{figure}
Fig.13 illustrates the convergence of the ergodic sum rate for the two-timescale MA design under MRT beamforming. The green curve represents the NLoS component modeled as a spatially correlated Rayleigh process with a correlation matrix, while the brown curve assumes an i.i.d. complex Gaussian NLoS component with an identity covariance matrix, constrained by the minimum antenna spacing. The small final gap observed for Rician factors $\kappa=$ 5, 10, and 15 dB validates that the minimum spacing constraint effectively mitigates spatial correlation, justifying the use of the simplified NLoS model for computational efficiency.
\begin{figure}[!t]
\centering
\includegraphics[width=2.4in,height=1.8in]{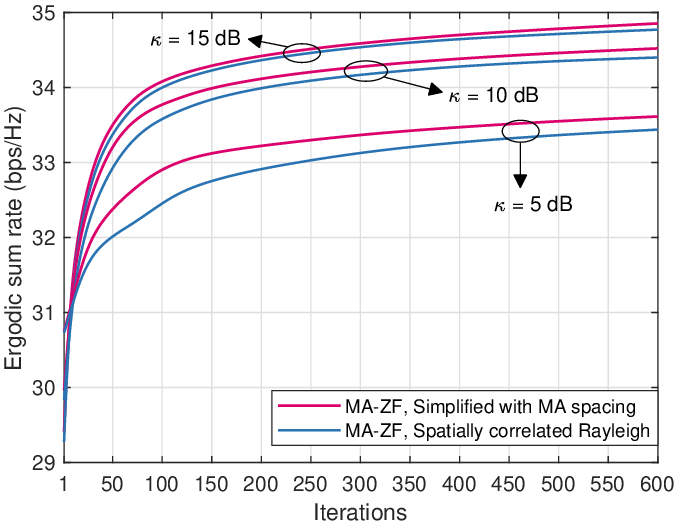}
\caption{Ergodic sum rate evaluation for MA-ZF with different NLoS Models ($N=8,M=5$).} 
\vspace{-9pt}
\end{figure}
\begin{figure}[!t]
\centering
\includegraphics[width=2.4in,height=1.8in]{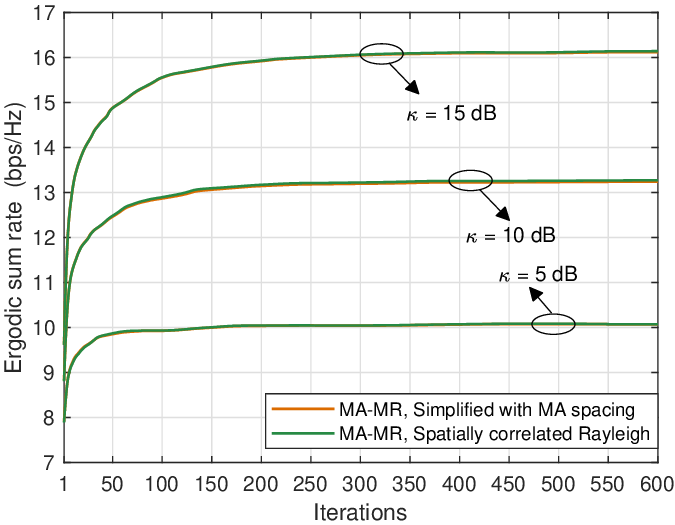}
\caption{Ergodic sum rate evaluation for MA-MRT with different NLoS Models ($N=6,M=5$).} 
\vspace{-9pt}
\end{figure}
\begin{figure}[!t]
\centering
\includegraphics[width=2.4in,height=1.8in]{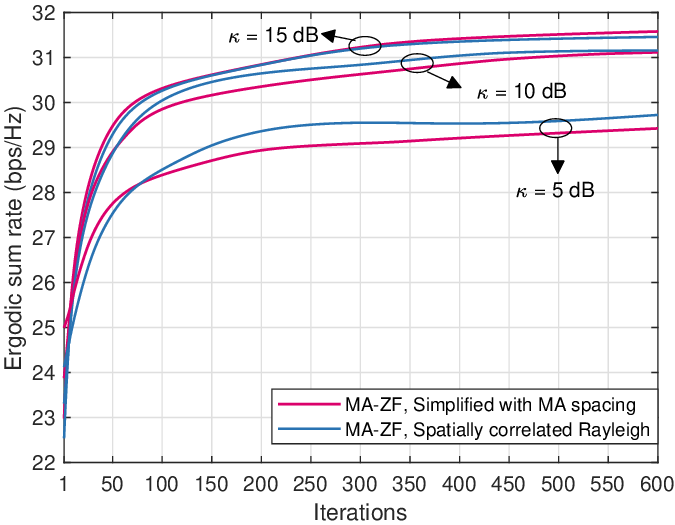}
\caption{Ergodic sum rate evaluation for MA-ZF with different NLoS Models ($N=6,M=5$).} 
\vspace{-9pt}
\end{figure}
Figs. 14 presents the ergodic sum rate evolution for an MA system with ZF beamforming. The blue curve accounts for spatial correlation in the NLoS component, while the red curve represents the i.i.d. assumption with an identity covariance matrix under the minimum spacing constraint. The small gap, within 1 bps/Hz, suggests that the i.i.d. model provides a reasonable approximation, particularly at higher Rician factors where the LoS component dominates.
Figs. 15 and 16 further investigate the impact of spatial correlation using fewer transmit antennas. The results indicate that spatial correlation can slightly degrade or improve performance depending on antenna configuration and repositioning. Compared to MRT, ZF beamforming is more sensitive to spatial correlation effects. Nonetheless, the numerical results remain consistent with the simplified model under the minimum spacing constraint.
Moreover, while spatial correlation effects intensify as the array size increases, the MA architecture inherently mitigates these effects. By leveraging spatial degrees of freedom without requiring additional antennas or RF chains, MAs provide a practical means to counteract correlation-induced performance variations.

Figs. 17 and 18 assess the sensitivity of the proposed two-timescale design to position estimation errors, evaluating the impact of inaccuracies in both user and antenna positions on the ergodic sum rate. User position errors are modeled as uniformly distributed within $[-1,1]$ meters, while the antenna position errors ranged from $\left[-\frac{\lambda}{20},\frac{\lambda}{20} \right]$.
\begin{figure}[!t]
\centering
\includegraphics[width=2.4in,height=1.8in]{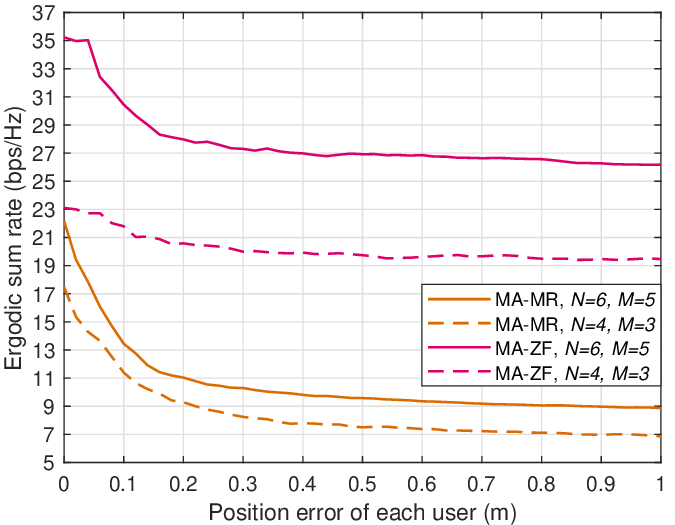}
\caption{Performance degradation with user position error with $\kappa=20dB$.} 
\vspace{-9pt}
\end{figure}
\begin{figure}[!t]
\centering
\includegraphics[width=2.4in,height=1.8in]{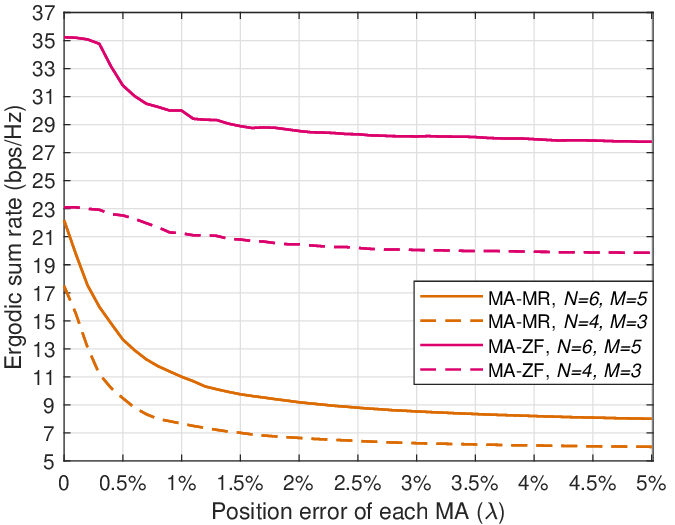}
\caption{Performance degradation with antenna position error with $\kappa=20dB$.} 
\vspace{-9pt}
\end{figure}
Figure 17 plots the ergodic sum rate as a function of user position errors (0 to 1 meter) under different configurations. The sharpest decline occurs within the first 0.1 meters, indicating high sensitivity to small errors. Figure 18 illustrates the degradation in ergodic sum rate due to antenna position errors (0\% to 5\% of $\lambda$), showing a similar sensitivity. The impact is particularly pronounced for the MA-MRT scheme, which relies heavily on precise antenna repositioning for optimal performance.
In summary, user position errors at the decimeter scale induce phase mismatches in the channel matrix, affecting the LoS signal alignment between the BS and the user. Likewise, MA positioning inaccuracies pose a significant challenge, as mechanical movement errors can cause subwavelength deviations from the optimized results, leading to performance degradation.
Despite these challenges, the numerical results suggest that our proposed optimization algorithms effectively identify local optima for the MA system, where small errors or deviations can cause significant performance loss\footnote{To accommodate user mobility and avoid user positioning error, frequent updates of statistical CSI and the corresponding MA positions adjustment become necessary. In practice, the MA functionality can be selectively disabled or updated less frequently depending on the deployment scenario, making the performance gain optional when precise positioning cannot be maintained. In addition, the micro-electromechanical systems (MEMS)-integrated antenna and the liquid antenna are two alternative solutions for implementing MAs with very high positioning accuracy (micro- to even nano-meter scale) in portable devices with compact components.}.

Figs. 19 and 20 compare the ergodic sum rate achieved by our surrogate-based algorithms against a near-optimal brute-force solution. The brute-force approach searches for optimal antenna positions over a discrete grid with a predefined number of points. We evaluate two configurations: (i) $N=4,M=3$ within $x\in[-1.6\lambda,1.6\lambda],y\in[-1.6\lambda,1.6\lambda]$ and (ii) $N=6,M=5$ within $x\in[-1.6\lambda,1.6\lambda],y\in[-1.6\lambda,2.4\lambda]$, both under $\kappa=20dB$. 
\begin{figure}[!t]
\centering
\includegraphics[width=2.4in,height=1.8in]{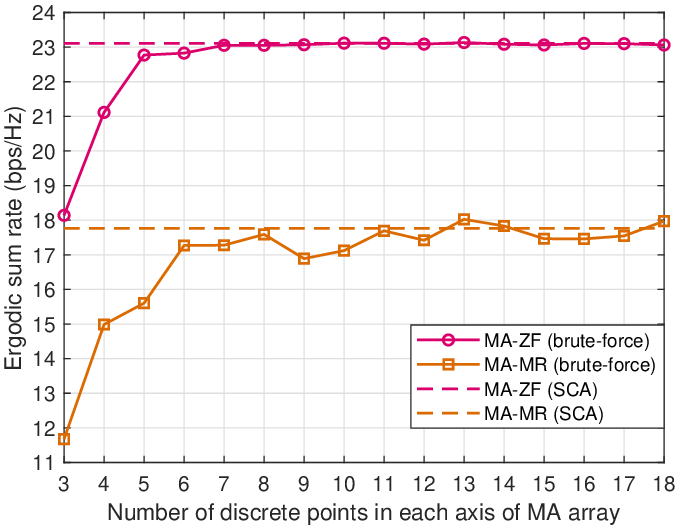}
\caption{Performance comparison with the brute-force method with $N=4,M=3$.} 
\vspace{-8pt}
\end{figure}
\begin{figure}[!t]
\centering
\includegraphics[width=2.4in,height=1.8in]{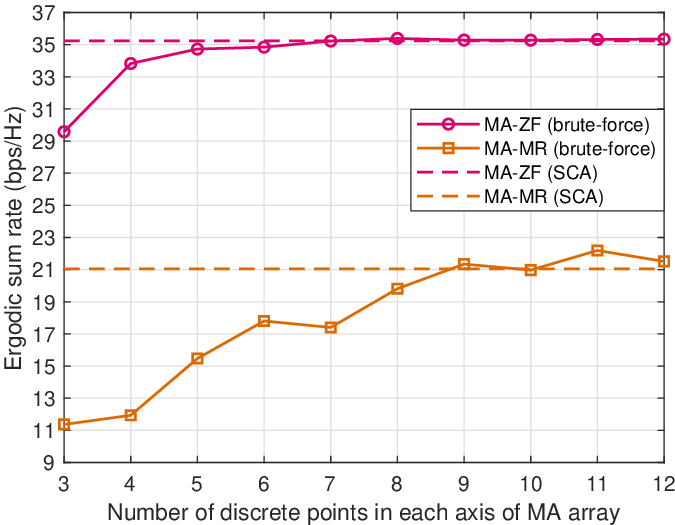}
\caption{Performance comparison with the brute-force method with $N=6,M=5$.} 
\vspace{-8pt}
\end{figure}
In Fig. 19, the dashed purple and orange lines represent brute-force results for MA-ZF and MA-MRT, respectively, while the solid lines correspond to our method. As grid resolution increases, the brute-force solution improves and stabilizes. Our method closely approximates brute-force performance, with only slight degradation. Notably, for an $N$-antenna MA array, the brute-force approach requires an $N$-dimensional exhaustive search over two spatial dimensions per antenna. With $D$ discrete points per axis, the total number of evaluations scales as $D^{2N}$, making it computationally prohibitive even for moderate $N$ (for example, $D=10, N=4$ requires $10^8$ evaluations. 
The trends in Fig. 20 mirror those in Fig. 19, with the surrogate-based method exhibiting a small performance gap within 6\%, demonstrating its near-optimality. 
Although our approach sacrifices a marginal degree of optimality, it significantly reduces complexity in polynomial time, compared to the brute-force method on exponential scale, which becomes infeasible for practical MA systems.

\vspace{-12pt}
\section{Conclusion}
\vspace{-3pt}
In this paper, we proposed a two-timescale transmission framework for MA-enabled MU-MIMO systems. The large-timescale optimization leverages statistical CSI to design optimal MA positions, maximizing long-term ergodic performance. In the small timescale, with MA positions fixed, MRT or ZF beamforming vectors are determined based on instantaneous CSI to adapt to short-term channel fluctuations. This decoupling of MA position optimization from the instantaneous transmission process provides a solution that strikes a balance between performance and practicality, effectively reducing the update frequency of MAs' positions and lowering channel estimation overhead.
Within this framework, we developed position optimization algorithms for both MRT and ZF beamforming schemes. For MA with MRT, we proposed an AO and SCA-based algorithm that iteratively optimizes antenna positions, thereby refining the approximation of the ergodic sum rate. Similarly, for MA with ZF beamforming, we used AO, SCA, and MM techniques to iteratively maximize the ergodic sum rate through a lower-bound surrogate objective function.
Extensive numerical results validated the effectiveness of the proposed two-timescale design, demonstrating significant gains in ergodic sum rate compared to conventional FPA systems. The results highlighted the superiority of MA with ZF beamforming, particularly in moderate LoS conditions and high user density. In contrast, MA with MRT beamforming offers a simplified alternative to more complex beamforming designs in strong LoS conditions with moderate user density. Moreover, these findings indicate the synergy of combining beamforming and MA techniques for effective interference management. 

Building upon our current two-timescale design, future research will explore hierarchical or multi-timescale joint beamforming and antenna position optimization in MA-enabled systems, integrating techniques such as predictive CSI and reinforcement learning (RL) to dynamically adapt to fast-fading channels. Specifically, by tailoring these adaptive optimizations to user densities, spatial distributions, and varying channel conditions, the system can retain the benefits of large-scale statistical optimization and small-scale beamforming while incorporating flexible medium-timescale adjustments to handle significant channel variations. These approaches hold the potential to enhance the performance and practicality of the MA technique, balancing dynamic repositioning gains against practical limitations such as mechanical latency and power consumption, and thus position MA-enabled MIMO systems as a key solution for diverse next-generation wireless networks.


\ifCLASSOPTIONcaptionsoff
  \newpage
\fi

\end{document}